\documentclass[twocolumn]{aastex61}

\usepackage{amsmath}

\newcommand{\vect}[1]{\mathbf{#1}}   

\received{...}
\revised{...}
\accepted{...}
\submitjournal{ApJ}

\shorttitle{Energetic particles at T~Tauri stars}
\shortauthors{...}

\begin{document}

\title{Mottled protoplanetary disk ionization by Magnetically-channeled T Tauri star energetic particles}

\correspondingauthor{Federico Fraschetti}
\email{ffrasche@lpl.arizona.edu}

\author{F. Fraschetti}
\affiliation{Depts. of Planetary Sciences and Astronomy, University of Arizona, Tucson, AZ, 85721, USA} 
\affiliation{Harvard Smithsonian Center for Astrophysics, Cambridge, MA, USA} 

\author{J. J. Drake}
\affiliation{Harvard Smithsonian Center for Astrophysics, Cambridge, MA, USA} 

\author{O. Cohen}
\affiliation{Lowell Center for Space Science and Technology, University of Massachusetts, Lowell, MA 01854, USA}

\author{C. Garraffo}
\affiliation{Harvard Smithsonian Center for Astrophysics, Cambridge, MA, USA}



\begin{abstract}
The evolution of protoplanetary disks is believed to be driven largely by angular momentum transport resulting from magnetized disk winds and turbulent viscosity. The ionization of the disk that is essential for these processes has been thought due to host star coronal X-rays but could also arise from energetic particles produced by coronal flares or by travelling shock waves and advected by the stellar wind. We have performed test-particle numerical simulations of energetic protons propagating into a realistic T~Tauri stellar wind, including a superposed small-scale magnetostatic turbulence. 
The isotropic (Kolmogorov power spectrum) turbulent component is synthesised along the individual particle trajectories. We have investigated the energy range $[0.1 - 10]$ GeV, consistent with expectations from {\it Chandra} X-ray observations of large flares on T~Tauri stars and with recent indications by the {\it Herschel} Space Observatory of a significant contribution of energetic particles to the disk ionization of young stars. In contrast with a previous theoretical study finding dominance of energetic particles over X-ray in the ionization throughout the disk, we find that the disk ionization is likely dominated by X-rays over much of its area except within narrow regions where particles are channeled onto the disk by the strongly-tangled and turbulent magnetic field. The radial thickness of such regions is $\sim 5$ stellar radii close to the star and broadens with increasing radial distance. This likely continues out to large distances from the star ($10$ AU or greater) where particles can be copiously advected and diffused by the turbulent wind.
\end{abstract}




\section{Introduction} \label{sec:intro}

Angular momentum transport within a protoplanetary disk governs the flow of material toward and away from the star and is one of the primary drivers of disk evolution affecting the formation and migration of planets within the disk \citep[e.g.][]{Bodenheimer:95,Coleman.Nelson:16,Bai:17}.  The important transport processes are thought to be the turbulence resulting from the magneto-rotational instability (MRI; \citealt{Balbus.Hawley:98}), stresses due to large-scale magnetic fields driving an outflow \citep{Wardle.Koenigl:93,Koenigl.etal:10,Bai.Stone:13,Bai:17} and shearing within the disk \citep{Turner.Sano:08}.
All these magnetic angular momentum transport processes require that the gas be sufficiently ionized to couple to magnetic fields. However, models of protoplanetary disks heated by the UV-IR spectral energy distributions of their host stars are too cold for significant thermal ionization, except in the very inner disk regions within about ten stellar radii \citep[e.g.][]{Gammie:96}.   

\citet{Gammie:96} originally appealed to the ionizing powers of cosmic rays to ionize the disk sufficiently to become MRI ``active'' in the outer layers, so that accretion proceeds more rapidly with increasing vertical height in a layered fashion.  More recent work has found that the MRI is suppressed by ambipolar diffusion in the inner disk region ($\leq 15$~AU; \citealt{Bai:13,Bai.Stone:13}) and that a layered accretion flow is more applicable to the outer rather than the inner disk.  Accretion is instead most likely driven by a magnetized disk wind \citep{Bai:14}.  \citet{Bai:17} has emphasised that the disk wind kinematics and the disk magnetic field geometry and evolution are key to understanding global protoplanetary disk evolution.  The ionization state of the disk gas remains a central ingredient for each of these problems.  Since numerical models of planet formation in a protoplanetary disk have demonstrated the importance of disk gas dynamics in controlling the types of planets that can be formed \citep[e.g.,][]{Coleman.Nelson:16}, disk ionization and its spatial and temporal variation then plays a vital role in both determining the radial structure of the disk, its dynamics and the nature of any planetary system that forms within it.

In either a disk wind or MRI dominated regime it now appears that cosmic ray ionization and driving of accretion runs into difficulties because they can be effectively shielded from the disk by the magnetized T~Tauri wind \citep{Cleeves.etal:13,Drake.etal:17}.  
Even in the absence of significant cosmic ray shielding,  Glassgold and co-workers \citep[e.g.][]{Glassgold.etal:97,Igea.Glassgold:99} realised that X-ray emission from the corona of a disk-hosting T~Tauri star will likely dominate the gas ionization in the zones of planet formation, out to 50~AU or more.  
The penetrating power of X-rays provides ionization in deeper disk layers than can be reached by UV and EUV photons.  Ionization by the central star is also pivotal for disk chemistry in intermediate and surface layers \citep[e.g.][]{Aikawa.Herbst:99,Semenov.etal:04,Walsh.etal:12} and could additionally be responsible for alteration of the structure of dust grains \citep[e.g.][]{Glauser.etal:09,Ciaravella.etal:16}.

While X-ray emission is now accepted as a major ionization source and angular momentum transport driver, \citet{Turner.Drake:09}, hereafter TD09, suggested that flare energetic particles---analogous to solar energetic particles (EPs) accelerated during large solar flares---can contribute significantly to disk ionization. The basis for the TD09 result was the estimate by \citet{Feigelson.etal:02} that energetic proton fluxes accelerated as a result of flaring in T~Tauri stars should be of the order of $10^5$ times those of the contemporary Sun. They noted that such fluxes were consistent with proton spallation being responsible for the observed meteoritic abundances of several important short-lived radioactive isotopes.  In support of this, evidence has been found using the {\it Herschel} Space Observatory of enhanced ionization of a protoplanetary disk due to energetic protons of at least GeV energies \citep{Ceccarelli.etal:14}, although the inferred particle flux is challengingly high at 4--5 orders of magnitude above the  \citet{Feigelson.etal:02} estimate. 

Using the proton flux estimate of \citet{Feigelson.etal:02}, TD09 found that stellar energetic protons have 40 times the ionizing power of coronal X-rays, from the innermost regions of the disk out to much larger distances (e.g., $\sim 10^4$ stellar radii or $100$ AU) and are the strongest ionization process for the protoplanetary disk they modelled.  The TD09 study has potentially important implications.  While X-ray fluxes can be directly observed from T~Tauri stars and readily included in models of protoplanetary disks \citep[e.g.,][]{Ercolano.etal:09,Mohanty.etal:13}, EPs cannot be observed directly.  If EPs dominate the ionization of the outer disk layers, current models attempting to predict the chemistry, ionization state, heating, cooling, winds and dynamics of disks---processes thought to be at least partially controlled by X-rays---will need to be revised.  

Being the first study of its kind, the TD09 work employed some simplifying assumptions.  The most important of these in the context of the present work is that the EPs travel in straight lines.  Solar energetic particles are known to follow the interplanetary (solar wind-borne) magnetic field to a large extent, though are able to traverse field lines in regions of significant MHD turbulence.  Since T~Tauri stars have strong surface magnetic fields and drive magnetized winds, their EPs will be expected to be trapped or deflected by the turbulent, magnetized T~Tauri wind.  The wind itself will also be deflected around the disk. \citet{Feigelson.etal:02} note that the meteoritic isotopic anomalies present direct evidence that EPs do indeed impact the disk.  However, it is quite uncertain what fraction of them, and under what acceleration circumstances and magnetic field characteristics, have trajectories favorable to disk penetration.

Given the potential importance of T~Tauri EPs for disk ionization and chemistry, further study and assessment is warranted. Ionization and its effects on chemistry due to stellar EPs within the disk, assumed to be a slab, has been modelled in \citet{Rab.etal:17} who also assumed a rectilinear transport of particles within the circumstellar environment.
Under the assumption that the EPs propagation is purely diffusive/advective, \citet{Rodgers-Lee.etal:17} solved the transport equation and found that EPs dominate the disk ionisation only beyond $0.3$ AU, finding that they dilute with the distance from the star, $r$, as $r^{-1}$, whereas stellar $X$-rays simply dilute as $r^{-2}$. 

In this paper, we examine the propagation of EPs in the environment of T~Tauri stars with no assumption on diffusive motion.  We adopt a magnetosphere, wind and extended magnetic field structure computed using a state-of-the-art MHD model commonly applied to the solar wind, and recently applied to study the winds and magnetospheres of other stars.  We apply a particle transport model honed to model SEP propagation within the solar system to the magnetized, turbulent medium of the model T~Tauri wind.  We examine the predicted ionization rate as a function of radial distance and compare the results with the baseline TD09 study.

\section{T TAURI WIND AND MAGNETOSPHERIC MODEL}
\label{s:windmod}

As our test case, we adopt one of the wind and magnetosphere models computed by \citet{Drake.etal:17} for examining the shielding of cosmic rays by the stellar magnetic field and wind.  We refer the reader to that work for further details and describe the model here only in brief. 

The T~Tauri wind model employed a recent version of the BATS-R-US Magnetohydrodynamics (MHD) code originally developed for the solar corona \citep{Powell.etal:99} as part of the Space Weather Modeling Framework \citep{Toth.etal:05,Toth:12}.  The model  is based on Alfv\'en wave dissipation and couples, in a single simulation, the coronal thermodynamics and the solar wind acceleration in a self-consistent manner \citep{Sokolov.etal:13,Oran.etal:13,vanderHolst:14}.  When applied to the solar case, the model employs high resolution magnetograms as a basis for the lower chromospheric magnetic field that drives the coronal heating and wind acceleration.  \citet{Drake.etal:17} used a surface magnetogram based on that constructed for the classical T~Tauri star V2129~Oph from Zeeman-Doppler imaging by \citet{Donati.etal:07}.  V2129~Oph is a typical T~Tauri star, with an age of about 2~Myr, a mass of $1.35 M_\odot$, a radius of $2.4 R_\odot$, a rotation period of 6.53~d, and an accretion  rate of $\sim 10^{-8}M_\odot$~yr$^{-1}$ \citep{Donati.etal:07}.  \citet{Drake.etal:17} employed different rotation periods in their models to examine its influence on cosmic ray shielding.  A faster rotation rate increases the azimuthal winding, and consequently the strength, of the magnetic field in the super-Alfv\'enic ``Parker spiral'' regime.  For this work, 
we adopt the model computed for a rotation period of $4$ days, mass $M_\star = 0.5 M_\odot$ and radius $R_\star = R_\odot$, consistent with observations \citep[e.g.][]{Johns-Krull:07}.  The model output comprised the steady-state magnetic field, together with the a complete description of the wind and its flow on a three-dimensional cartesian grid. The simulation box is taken to be a star-centered cube with volume $(48 R_{\star})^3$, which is sufficiently large to enclose the stellar ``Alfv\'en surface'' representing the point at which the wind becomes supersonic. The consequent simulation cell size is $0.36 R_\star$.

\section{Energetic particles in the circum-T Tauri environment}\label{environ}

Solar wind {\it in-situ} measurements generally exhibit an enhancement in the intensity of EPs corresponding to the passage of interplanetary shocks, with a broad variety of behavior  such as enhancements delayed by a few hours, spikes, or no features at all \citep{Lario.etal:03}. Such measurements lead to the conclusion that shocks are efficient sources of accelerated particles. In the vicinity of T Tauri stars, intense flaring and eruptive activity is likely to produce copious shocks propagating in front of coronal mass ejections (CMEs) emanating from the star. In addition, if the Sun serves as a guide, such shocks are likely to be strong thereby producing hard particle spectra \citep[e.g.]{vanNes.etal.84} that can provide a source of ionization for the protoplanetary disk. An alternative process that can efficiently produce non-thermal particles (electrons) in the proximity of the Sun is generally accepted to be the solar flare, as confirmed by hard $X$-rays due to bremsstrahlung emission of accelerated electrons \citep{Holman.etal:11} and by microwave radiation due to gyrosynchrotron of trans-relativistic electrons spiralling in the coronal magnetic field \citep{White.etal:11}. In our simulations, we calculate the propagation of EPs only, allowing the aforementioned processes of acceleration to operate in tandem and to produce $\sim$ GeV protons at a variety of distances from the star; only the position of particle injection into the circumstellar medium ($R_s$) is prescribed.

Our simulations explore two separate {characteristics} of disk ionization by EPs: 1) the topology of the stellar large-scale magnetic field wrapped around the T Tauri star; 2) turbulent fluctuations on small scales. {Both combined} will provide a realistic picture of the propagation of the EPs in the circumstellar medium of a T Tauri star. 

\subsection{General considerations on the effect of turbulence}\label{turbulence}

\begin{figure}
	\includegraphics[width=9.cm]{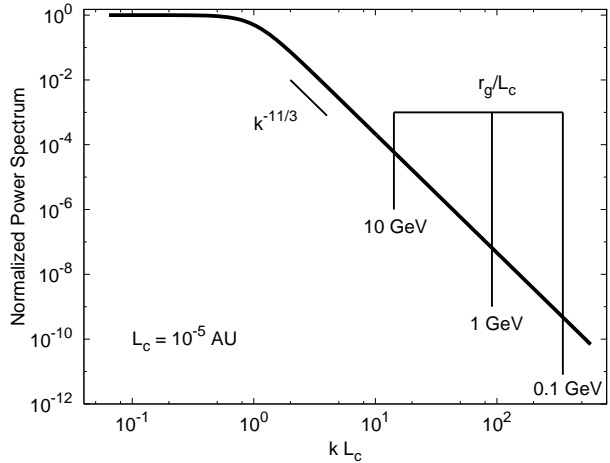}
\caption{Power-spectrum of the magnetic turbulence used in the test-particle numerical simulations. The vertical lines correspond to the resonant wavenumbers in the magnetic field at a radius $R_s = 5 R_{\star}$ ($B_0 \simeq 0.55$~G) sampled by individual protons with  energies $E_k=0.1, 1,10$~GeV in turbulence with $L_c = 10^{-5}$~AU.  \label{Pspectrum}}
\end{figure}

Observations of interplanetary magnetic turbulence \citep[e.g.]{Jokipii.Coleman:68}, and interstellar density measurements \citep{Armstrong.etal:95} indicate that turbulent fluctuations are typically distributed as a power-law in the turbulent wavenumber (see Fig.~\ref{Pspectrum}). The small-scale turbulence leads to pitch-angle scattering and cross-field motion of particles, both of which are relevant in general to the propagation of charged particles through a turbulent plasma. As a consequence, the EP trajectories that we calculate significantly differ from radial, as originally assumed by TD09.

The amount of scattering of EPs depends on the level of turbulence in the magnetic field and on the possible amplification of plasma waves by EPs streaming along the field; the latter contribution is neglected in this work. In our simulations, the EPs are injected at various distances from the central star into the circumstellar medium and spend most of their time before hitting the disk within a radial distance $R < 10 R_{\star}$ (here we chose $R_{\star} = R_\odot = 6.957 \times 10^5 ~{\rm km} = 4.6 \times 10^{-3} $ AU). 

Solar wind {\it in-situ} measurements at a variety of helio-latitudes by {\it Helios, IMP 8} and {\it Ulysses} \citep[see Fig. 4 therein]{Horbury.etal:96} have shown a decrease of the correlation length of the turbulence, $L_c$, i.e., the scale of the injection of the turbulence (see Fig. \ref{Pspectrum}), between a distance from the Sun of $\sim 1$ AU ($L_c \simeq 0.01$~AU) sun-ward down to $\sim 0.1$~AU ($L_c < 0.001$~AU). In the absence of current observational estimates of $L_c$ within a T~Tauri wind close to the star, at distances $R < 10 R_{\star} < 0.1$ AU the range $L_c = 10^{-6}$--$10^{-4}$~AU  (or $2.2 \times 10^{-4}$--$2.2 \times 10^{-2} R_\star$) seems to be a reasonable approximation. Based on solar wind observations, that represent one single realization of the unperturbed field and of the turbulence, it is hard to identify a correlation between $L_c$ and the field magnitude $B_0$.  It is then not obvious that the strong field of a T~Tauri star would lead to a different $L_c$. On the other hand, our analysis confirms that the choice of $L_c$ does not have a large effect on the statistical properties of EP propagation or the distribution of the hitting-points on the disk (see Appendix \ref{s:Lc}). Likewise, it is possible that $L_c$ depends to some extent on the stellar rotation velocity, though by the same argument we do not expect the short rotation period ($4$ days) as compared to the Sun would significantly affect the results.

Since EPs in our simulations sample a relatively small region ($\sim 0.1$ AU), we assume that $L_c$ is uniform within the simulation box; in future work this condition will be relaxed. The chosen $L_c$-values ensure that the EPs are resonant within the turbulent inertial range (see Fig. \ref{Pspectrum}) during their propagation, namely 
\begin{equation}
k_{\rm min} r_g({\mathbf x})/2\pi = r_g({\mathbf x})/L_c < 1, 
\end{equation}
where $r_g ({\mathbf x}) = p_\perp c/ e B_0 ({\mathbf x})$ is the gyroradius for an EP with momentum (or velocity) perpendicular to the unperturbed and space-dependent magnetic field of the T~Tauri star, $B_0 ({\mathbf x})$, given by $p_\perp = m v_\perp \gamma$ (or $v_\perp$), $m$ is the EP mass, the Lorentz factor is $\gamma = 1/ \sqrt{1-(v/c)^2}$ with $c$ being the speed of light in vacuum. 

The parallel diffusion coefficient, $\kappa_\parallel$, in turbulence with the assumed power-spectrum can be calculated using quasi-linear theory \citep{Jokipii:66,Earl:74,Giacalone.Jokipii:99}; the resulting space-dependent mean free path $\lambda_\parallel ({\mathbf x}) = 3 \kappa_\parallel ({\mathbf x}) / v$ reads
\begin{equation}
\lambda_\parallel ({\mathbf x}) \simeq 4.8 (r_g({\mathbf x})/L_c)^{1/3} L_c/\sigma^2
\label{lambda}
\end{equation}
up to mildly relativistic (i.e., a few GeV) energy protons, where 
\begin{equation}
\sigma^2 = (\delta B ({\mathbf x}) /B_0 ({\mathbf x}))^2
\label{sigma}
\end{equation}
is the power of the space-dependent fluctuation $\delta B$ relative to $B_0$. In our simulations, we will assume $\sigma^2$ independent of space throughout the simulation box (see Appendix \ref{s:sigma} for the physical justification). 
Thus, $\lambda_\parallel ({\mathbf x})$ drops in the proximity of the T Tauri star as $\lambda_\parallel ({\mathbf x}) \propto L_c^{2/3} / B_0^{1/3}({\mathbf x})$ due to the increase of $B_0 ({\mathbf x})$; in a more realistic approach, such a drop is even faster due to the decrease of $L_c$ not accounted for in this work. For 1~GeV protons and $\sigma^2 = 0.1$, $\lambda_\parallel \ll 10 R_{\star}$; this value is further reduced if $\sigma^2$ approaches unity. Typically $\lambda_\parallel$ is smaller than the grid-cell size ($0.36 R_\star$), ensuring a sufficient number of scatterings within a single cell.

In addition to scattering, evidence has built up from solar wind observations that perpendicular diffusion leads to a significant angular spread of EPs up to $0.1$ GeV, both in longitudinal and latitudinal directions of the heliosphere \cite[and references therein]{Strauss.etal:17}. On the other hand, longitudinal tracking of Ground Level Enhancement events, i.e. SEP events with typical particle energy $\gtrsim 1$ GeV, is not available for solar cycles prior to the current one ($24^{th}$) and in the current one an unusual lack of events (only $2$ so far) prevents statistically significant conclusions. We refer to a two-fold type of perpendicular diffusion (see, e.g., \cite{Fraschetti.Jokipii:11}). A first contribution is due to the meandering of field lines described by the diffusion coefficient $\kappa_\perp^{MFL}$, i.e., the fluctuation of the magnetic field in the direction perpendicular to the local average field (this contribution assumes that the particle is simply chained to the magnetic field lines and is dominated by large-scale fluctuations).  The second contribution is the gradient/curvature drift diffusion ($\kappa_\perp^{D}$), due to the spatial variation of the turbulence $\delta B ({\bf x})$ bringing about a departure of the particle from the field line it was initially on. 

In general, the perpendicular diffusion coefficient grows with the particle energy and with the turbulence strength; for instance, in 3D isotropic turbulence $\kappa_\perp^{D}$ scales as $\kappa_\perp^{D} \propto (\sigma r_g)^2$ \citep{Fraschetti.Jokipii:11}, leading to a steep scaling with respect to the unperturbed field $\kappa_\perp^{D} \propto B_0^{-2}$, as $\sigma^2$ is approximately uniform. T~Tauri stars harbor kilogauss surface-averaged magnetic fields \citep[e.g.][]{Johns-Krull:07}, three orders of magnitude greater than the present day Sun. The aforementioned scaling of $\kappa_\perp^{D}$ with $B_0$ hampers the transport of EPs in the perpendicular direction within the circumstellar ambient medium of a T~Tauri star with respect to the Sun. Thus, close to the T~Tauri star the departure of EPs from field lines is expected to be very small whereas at larger distances EPs are more and more unleashed from the magnetic field. 

For a solar wind proton with a kinetic energy of a few GeV measured {\it in situ} at $1$ AU, and for a Parker spiral field $B_0 = 5 \times 10^{-5}$ G, the gyroradius is $r_g \sim 10^6$~km; such a gyroscale shrinks down to $ \sim 10^2$~km at a distance less than $10 R_{\star}$ from the T~Tauri star that we consider in our simulations. Therefore, $r_g$ is resonant with much smaller turbulent scales than in the solar wind at $1$ AU.  
Moreover, the turbulence in the circumstellar medium of a T~Tauri star is likely to be much stronger than in the solar wind, due to the much more violent stellar activity. In our simulations we therefore considered values of $\sigma^2 $ in the range $0.01 -1.0$, to be compared with the steady solar wind at 1~AU where typically $\sigma^2 = 0.01$--$0.03$ \citep{Burlaga.Turner:76}. 

Since both $\kappa_\perp^{MFL}$ and $\kappa_\perp^{D}$ grow linearly in $\sigma^2$, we might expect a greater contribution to latitudinal and longitudinal spread due to cross-field diffusion than in the solar wind at a given $B_0$ \citep{Dresing.etal:12,Fraschetti:16a,Fraschetti:16b}. However, such an effect is outweighed by the steep scaling of $\kappa_\perp^{D}$ with $B_0$ in the T~Tauri field, mentioned above. 

\subsection{Basic assumptions of the model}\label{assumptions}

We assume that the stationary MHD wind solutions are a good approximation for the purposes of our study as both the large-scale magnetic field and the turbulence are stationary on the time-scale of EP propagation. The justification is that the typical EP propagation speed ($\simeq c$) greatly exceeds both the stellar rotation speed close the surface (approximately 25~km~s$^{-1}$ for a rotation period of 4 days), 
and the Alfv\'en wave speed in the circumstellar medium (for $B_0 = 10$ G and $n=5\times 10^4$ cm$^{-3}$ at a distance $\sim 2R_{\star}$ from the star, the Alfv\'en speed is $\sim 2 \times10^4$ km$/$s and decreases outward as $\sim 1/R$). Thus, on the time-scale of EP propagation, the frame rotating with the star and the rest frame of the expanding plasma are, to good approximation, indistinguishable. As a consequence, it is reasonable to assume that the global EP anisotropy in the expanding wind frame is small. Thus, the wind advection has no effect on the EPs motion outward that is mainly of diffusive nature.

\citet{Feigelson.etal:02} used observations of intense flaring in Orion T~Tauri stars combined with a relation between X-ray and EP emission on the Sun to estimate that the circumstellar medium of a T~Tauri star will be enriched in EPs by a factor of $\sim 10^5$ compared with the contemporary Sun. The sources of EPs closer to the central star are likely to be shocks preceding CMEs or flares in the stellar corona. Here, we are interested in the space-averaged effect of EPs on disk ionization. We mimic such sources by uniformly injecting EP populations on spheres at various radii (from 2 to $10 R_{\star}$) with an isotropic velocity distribution.  {Our assumption of a stationary magnetic field distribution implicitly ignores the effects on the disk of transients.  Individual propagating shocks due to CMEs travelling across circumstellar space would disrupt the pre-existing large-scale magnetic field, resulting in changes in field topology and possibly in reconnection of field lines.} 

The rate of energy change, $\Delta E$, for a nearly isotropic population of EPs in a flow with speed $V$ is determined to the lowest order in $V/c$ by the divergence of the flow speed and can be crudely estimated in the plasma frame by 
\begin{equation}
\Delta E/E = (2V/3r) \Delta t 
\end{equation}
during the elapsed time $\Delta t$ \citep{Kota:13}. In a stationary wind with speed $V \simeq 800$ km$/$s, an appreciable $\Delta E / E$ occurs only over a $\sim 1$ day time-scale. In our simulations, the vast majority of the EPs 
hit the disk or return to the stellar surface within 1--2 hours for $\sigma^2=0.1$. Over such a short time-scale as compared with the dynamical time-scale of the advected bulk of the star wind, the energy of the EPs does not change appreciably and the adiabatic approximation is justified. In stronger turbulence ($\sigma^2 \gtrsim 1$), or at smaller particle energies, $\Delta t$ is greater as particles spend more time within the turbulent region ($\Delta t \propto \kappa_\parallel^{-1}$, where $\kappa_\parallel$ grows with energy and decreases with $\delta B$, regardless of the assumed turbulence model) and stellar modulation effects need to be accounted for. We neglect also the energy loss due to curvature drifts of particles moving against the motional electric field ${\bf V}/c \times {\bf B}$. 

We use simplifying assumptions for the radial and vertical structure of the protoplanetary disk. The total vertical column density typically decreases with radius to some power depending on the viscosity \citep[e.g.][]{Pringle:81,D'Alessio.etal:98}, while the scale height increases due to the weakening vertical component of stellar gravity with radius. In our simulations, we are primarily interested in EP trajectories that intersect the disk; thus, any disk radial and vertical structure is {neglected}. The disk lies on the XY-plane and is assumed to have semi-thickness $0.1 R_{\star}$ independent of distance ($< 30 R_{\star}$ from the star) and an innermost radius of $2 R_{\star}$. The actual disk structure will depend on its temperature stratification and the equilibrium between gas, magnetic pressure and wind ram pressures at the disk--magnetosphere boundary, and cannot be calculated self-consistently in our simulations that do not include heating or the wind and magnetic field of the disk. Thus, the MHD instabilities generated at the disk--magnetosphere boundary \citep{Kulkarni.Romanova:08} are not reproduced here (see Sect.~\ref{discuss}). In Appendix~\ref{s:disk} we show that a thicker disk or a greater disk innermost radius do not qualitatively change our results. 
At much greater distances from the star, the vertical structure needs to be considered in more detail; we defer this more complex treatment to future work. 

EPs are propagated within the circumstellar environment until their trajectories either intersect with the disk surface, or take them back to the stellar surface. The magnetic field of a T Tauri star is strongly curved and can be azimuthally wrapped, trapping EPs close to the star. As a consequence, the wind advection cannot efficiently drag EPs away from the star: the fraction of particles collapsing back onto the star, that in the stretched out Parker spiral field of the Sun is negligible, is significant and clearly larger for smaller injection radius, $R_s$. The fraction of particles escaping the simulation box is $< 10 \%$ for $R_s = 2 R_{\star}$, and grows up to $\sim 30 \%$ for $R_s = 8 R_{\star}$. Since the path-length to stop an EP is much smaller than the thickness of the disk, once a particle hits the surface of the disk we assume that its kinetic energy is converted instantaneously into ionization of the disk gas.

\section{Numerical Method}\label{nummeth}

In a series of numerical experiments, we consider a population of energetic charged test-particles (protons) gyrating in a turbulent magnetic field described as follows. We assume a three-dimensional magnetic field of the form 
\begin{equation}
{\bf B(x) = B}_0 ({\bf x}) + \delta {\bf B(x)}, 
\end{equation}
where the large-scale component, ${\bf B}_0 ({\bf x})$, is the 3D magnetic field generated by the MHD simulations as calculated in \citet{Drake.etal:17}, and the random component ${\bf \delta B} = {\bf \delta B} (x, y, z)$ has a zero mean ($\langle \delta {\bf B(x)} \rangle = 0$)  and a turbulence correlation length $L_c$. We assume an inertial range $k_{\rm min} < k < k_{\rm max}$, with $k_{\rm max}/k_{\rm min} = 10^2$, where $k_{\rm min} = 2\pi/L_c$ and $k_{\rm max}$ is the magnitude of the wavenumber corresponding to some dissipation scale marking the smallest scale of applicability of the ideal MHD. A broader inertial range does not substantially change the results presented in the following section, despite being computationally more expensive. The turbulence power spectrum (Fig.~\ref{Pspectrum}) is assumed to be scale-invariant, or Kolmogorov, in all the three space-dimensions: 
\begin{equation}
G(k) \propto k^{-\beta -2 }, 
\end{equation}
where $\beta = 5/3$ is the one-dimensional power-law Kolmogorov index and the additional $2$ accounts for the dimensionality of the turbulence. At scales larger than $k_{\rm min} ^{-1}$ ($k_{0} < k < k_{\rm min}$, with $k_{\rm min}/k_{0} = 10^2$ ), the power spectrum is simply taken as constant consistently with the solar wind large-scale power spectrum (see, e.g., \citealt{Jokipii.Coleman:68}).

The method of particle propagation follows the widely used approach in \cite{Giacalone.Jokipii:99} and \citet{Fraschetti.Giacalone:12}. The fluctuating field comprises $N_m$ transverse wave modes at every point occupied by the particle with random amplitude, phase, orientation and polarization. 
The magnitudes of the wavenumber $k_i$ of the $N_m$ modes are logarithmically equispaced, $\Delta k/ k =$~constant. We sampled the power spectrum over a discrete number of $N_m^* = 188$ modes. 
Numerical calculations show that the diffusion coefficients are not affected by the sampling resolution in $k-$space with $N_m > N_m^*$ \citep{Fraschetti.Giacalone:12}. However, the fluctuation $\delta {\bf B}$ is calculated here with a space-dependent amplitude (see Appendix \ref{s:sigma}).

Test-particles are initialized in the fully twisted magnetic field and evolved according to the Lorentz force determined by the total magnetic field at the instantaneous particle position. The equation of motion numerically solved for particles with charge $e$ and mass $m$ moving with velocity $\vect{v} (t)$ in a magnetic field $\vect{B}(\vect{x})$ is the Lorentz equation
\begin{equation}
\frac{d\vect{u}(t)}{dt} =  \vect{u}(t)
\times\vect{\Omega}(\vect{x})\;,
\label{lorentz}
\end{equation}
where 
\begin{equation}
\vect{\Omega}(\vect{x})  = e \vect{B}(\vect{x})/(mc\gamma)
\end{equation}
with $\vect{u} (t) = \gamma \vect{v} (t)/c$ at the time $t$.
The quantity $\vect{\Omega}(\vect{x})$ in Eq.(\ref{lorentz}) is given by 
\begin{equation}
\vect{\Omega}(\vect{x}) = \vect{\Omega}_0 (\vect{x})+\delta \vect{\Omega}(\vect{x})
\end{equation}
where 
\begin{equation}
\vect{\Omega}_0 (\vect{x}) \equiv {e\over mc\gamma} \vect{B}_0 (\vect{x})   , \quad   \delta \vect{\Omega}(\vect{x}) \equiv {e\over mc\gamma} \delta \vect{B} (\vect{x}).
\end{equation}
in terms of the background magnetic field $\vect{B}_0$ and of $\delta \vect{B}$; we adopt a single realization of $\delta {\bf B}$ (see Appendix \ref{s:sigma}).   

The equation of motion \ref{lorentz} is written in the local frame of the expanding plasma: the magnetic field transforms in a fluid in motion with velocity $\vect V$ as ${\vect B}_\parallel = {\vect B}'_\parallel$ and ${\vect B}_\perp = \Gamma({\vect B}'_\perp - {\vect V}/c \times {\vect E}')$ in the directions parallel (${\vect B}_\parallel$) and perpendicular (${\vect B}_\perp$) to ${\bf V}$, respectively, with $\Gamma = 1/ \sqrt{1-(V/c)^2}$. For a T~Tauri wind $V/c \simeq 0.003$, $\Gamma \simeq 1$, and ${\vect B} \simeq {\vect B}'$. The plasma is assumed to be infinitely conductive so that any large-scale background electric fields are negligible; also the rate of energy change is so small that any motional electric field associated with the turbulent field is neglected. We also neglect the electric field associated with the gradient/curvature drift. In order to calculate $\vect{B} ({\bf x})$ at the particle position, we do not perform a magnetic field interpolation within the cell of the MHD grid, but simply calculate the field in the cell closest to the particle position; using the former would be more time-consuming as all six values of the B-components per cell-grid need to be computed, one on each face, with little advantage in numerical accuracy. 

Equation (\ref{lorentz}) is integrated by using two encapsulated time-steps: a coarse time-step $\sim 65 \times 2\pi/\Omega_0({\bf x})$ that varies according to the particle energy and $\vect B_0$, i.e., with the distance from the star; and a fine time-step for which we use the Bulirsch-Stoer method with adjustable time-step \citep{Press.etal:86}. This allows us to efficiently deal with variations of $B_0$ by orders of magnitude along the EPs propagation. We verified that the particle energy is conserved to a relative accuracy of $\sim 10^{-3}$.

The computations presented here are limited in two ways: by the size of our stellar wind MHD simulation box which limits the radial extent of the disk investigated to $24R_\star$; and by the test-particle approach which constrains the number of particles we can treat due to computational time limitations.

\begin{figure}
	\includegraphics[width=10cm]{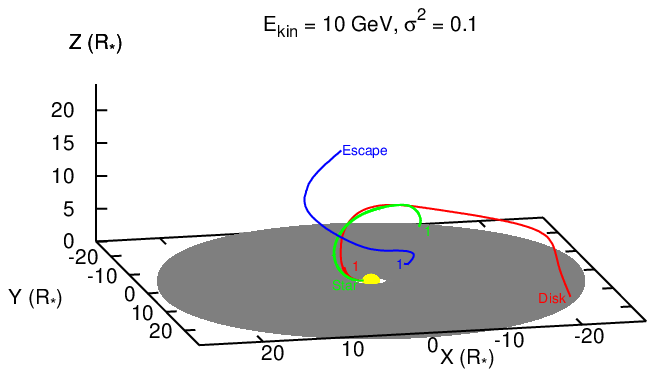}\\
	\includegraphics[width=10cm]{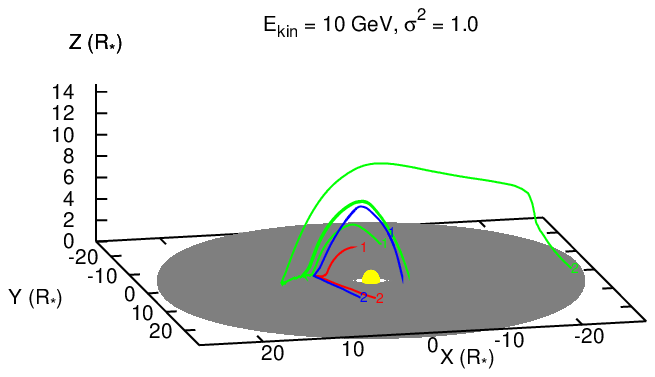}
\caption{{\it Upper} : Three-dimensional trajectories of three selected protons injected at $R_s = 5 R_{\star}$ with $E_{kin} = 10$~GeV and $\sigma^2 = 0.1$. The yellow hemisphere in the centre represents the star and the gray circle the inner portion of the disk. The ``1'' labels indicate the injection points for each particle, whereas the final point is labelled by the particle final fate: hitting the disk (red), collapsing onto the star (green) and escaping the box (blue). All axes are in $R_{\star} $ units. {\it Lower}: Three-dimensional trajectories of three selected protons injected at $R_s = 5 R_{\star}$ with $E_{kin} = 10$~GeV and $\sigma^2 = 1.0$. The injection point for each particle is labelled by ``1'', the final point by ``2''. All three particles hit the disk.   \label{trajectory}}
\end{figure}

\begin{figure}
	\includegraphics[width=12cm]{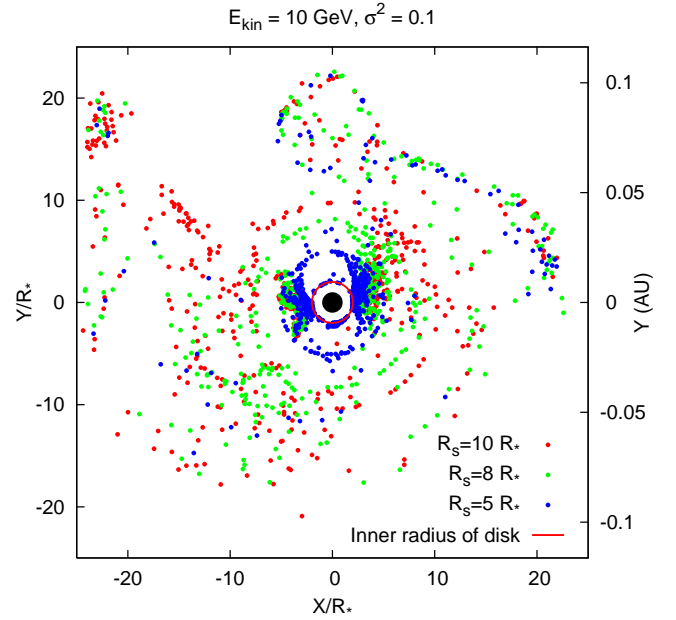}
\caption{Coordinates of the hitting points for 10~GeV kinetic energy protons on the disk plane for three selected values of injection distance on a sphere, radius $R_s= 5, 8$ and $10 R_{\star}$; here, $L_c = 10^{-5}$~AU and $\sigma^2 = 0.1$. The red circle with radius $2 R_{\star}$ marks the innermost radius of the disk. The $x$ and left $y$ axes are in units of $R_{\star}$, while the right $y$ axis is in AU.  \label{hit_source}}
\end{figure}

\section{Results}\label{results}

Figure \ref{trajectory}, upper panel, shows the trajectories of three particles injected into the circumstellar medium at $R_s = 5 R_{\star}$ with $\sigma^2 = 0.1$ having three different fates: hitting the disk, collapsing onto the star and escaping the simulation box. The lower panel instead shows three cases of EPs that hit the disk in a medium with $\sigma^2 = 1.0$.

Fig.~\ref{hit_source} illustrates the coordinates of the hitting points on the disk plane (the XY plane) for {various injection radii $R_s$ with $\sigma^2 = 0.1$ and $L_c = 10^{-5}$ AU}. 
For weak turbulence, the distribution of hitting points spreads over the XY plane mostly as a result of a larger injection radius: {particles injected at $R_s =10 R_{\star}$ (red dots) are spread on the plane more than those injected at $R_s =5 R_{\star}$ (blue dots)}; particles injected further out have lower chances of moving back toward the star and higher chances of hitting the disk. The distribution of disk hitting points is clearly non-uniform and takes on a mottled appearance, as particles are channeled by the ambient magnetic field into streams or sheets.

The remaining particles leave the simulation box and thus their fate is not determined by our simulations: they might contribute to the ionization in outer disk regions or, if they {scatter and travel back along the same field line or} switch to another field line as a result of perpendicular transport, either travel back toward the star or out of the circumstellar region without hitting the disk.

\begin{figure}
	\includegraphics[width=1\columnwidth]{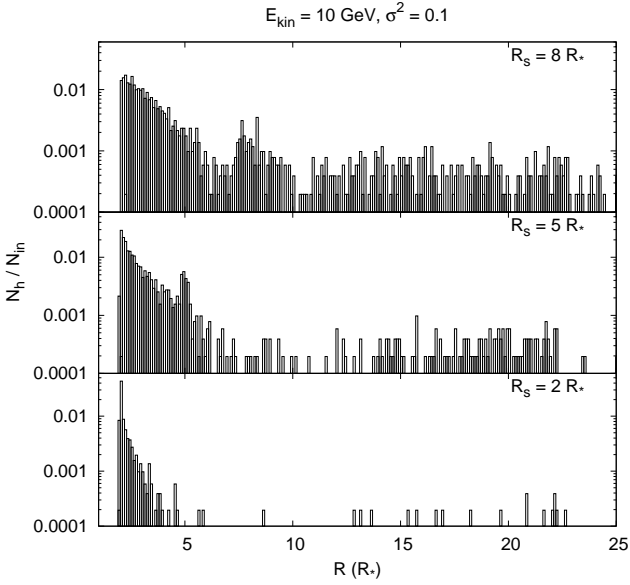}
\caption{Histogram of the ratio of the number of disk-hitting $10$ GeV particles in $0.1R_\star$ bins, $N_h$, to the total number of injected particles $N_{in}$ as a function of radial distance from the star, in units of $R_{\star} $
, for three values of injection distance on a sphere with radius $R_s= 2, 5, 8 R_{\star}$; here $\sigma^2 = 0.1$ and $L_c = 10^{-5}$ AU. Note that the y-axis is in log-scale. \label{histo_comp_se}}%
\end{figure}

The radial distributions of particle hitting-points on the disk for selected values of $R_s$ are shown in Fig. \ref{histo_comp_se}.  Here, the $y$-axis is the ratio of the number of disk-hitting particles integrated over radial bins of size $0.1 R_\star$, $N_h$,
to the total number of injected particles in the simulation, $N_{in}$. 
These all peak at the innermost disk radius $R_s = 2 R_{\star}$ regardless of the value of $ R_s$; in Appendix \ref{s:disk} we explored the distribution of hitting points for different innermost disk radius. Such a peak is dominated by the particles that are injected at intermediate co-latitudes $\vartheta$ 
({where $\vartheta = 90^\circ$ at the disk plane and $ 0^\circ$ at the pole})
on field lines magnetically connected to the inner part of the disk that are azimuthally wrapped around the star.  

\begin{figure}
	\includegraphics[width=1\columnwidth]{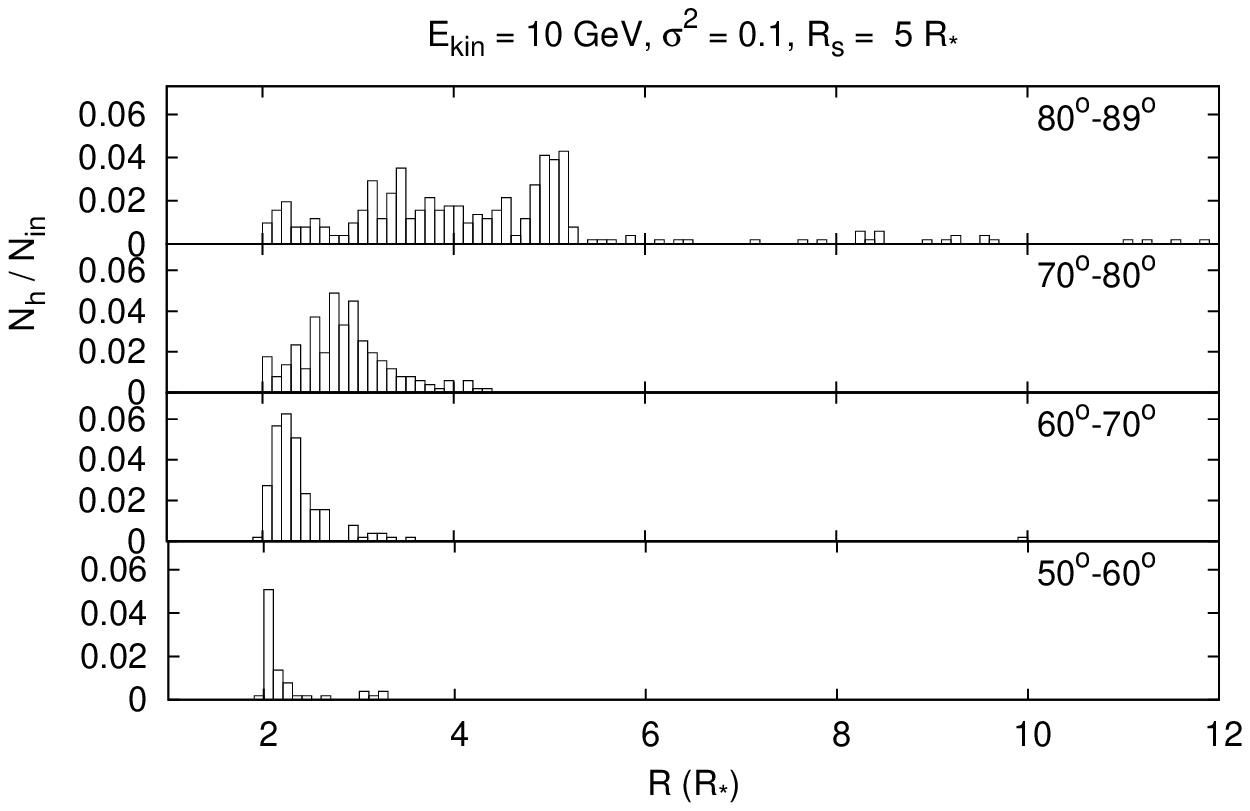}
\caption{Histogram of $N_h/N_{in}$ for $10$ GeV particles as a function of radial distance from the star (in units of $R_{\star} $) for $\sigma^2 = 0.1$ and $Lc=10^{-5}$ AU. EPs are injected within various rings of co-latitudine $\vartheta$ (from top to bottom $[80^\circ - 89^\circ]$, $[70^\circ - 80^\circ]$, $[60^\circ - 70^\circ]$, $[50^\circ - 60^\circ]$) on a spherical surface with radius $R_s= 5 R_{\star}$.  
\label{histo_comp_se_lat}}%
\end{figure}

The dependence of particle trajectory on injection co-latitude is further examined in Figure~\ref{histo_comp_se_lat} that shows, for $\sigma^2 = 0.1$, the hitting points of equal numbers of 10~GeV EPs injected on the sphere with radius $R = 5 R_{\star}$ within four adjacent co-latitudinal rings in the hemisphere $z>0$: mid- ($50^\circ - 60^\circ$) mid-high ($60^\circ - 70^\circ$ and $70^\circ - 80^\circ$) and high co-latitudes $\vartheta$ ($80^\circ - 89^\circ$). From high to mid-co-latitudes, the smaller the injection $\vartheta$ (i.e., the closer to the pole), the closer in the final hitting point is. EPs injected at mid-latitudes ($\vartheta$ $[50^\circ - 60^\circ]$, lowest panel) tend to follow the large-scale dipolar topology of the field lines and impact the disk close to the star.   Particles injected at lower co-latitudes, closer to the magnetic pole, are not reported in this figure as they collapse in large part onto the polar region of the star, and hence the larger number of particles ``lost'' to the $50^\circ$--$60^\circ$ histogram in the lower panel compared to the others.

\begin{figure}
	\includegraphics[width=12cm]{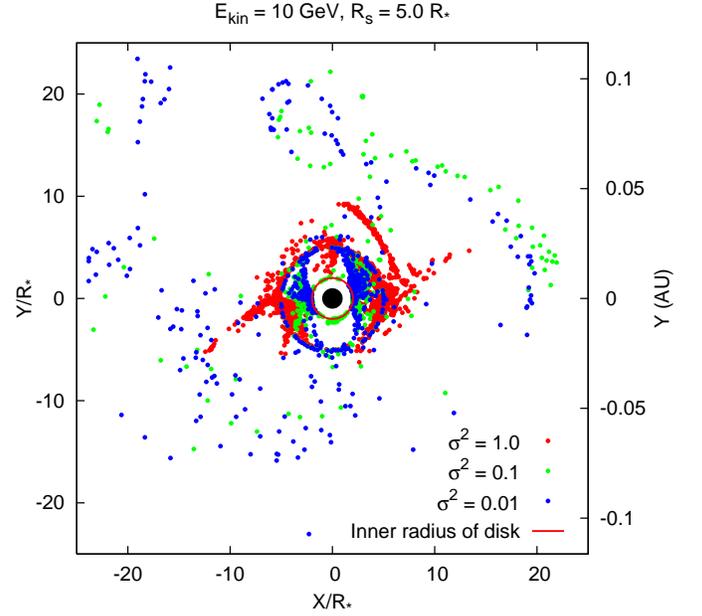}
\caption{Coordinates of the hitting-point for $10$ GeV kinetic energy protons on the disk plane for $\sigma^2 = 0.01$, $0.1$ and $1.0$. Protons are injected on a sphere with radius $R_s= 5 R_{\star}$; here $L_c = 10^{-5}$ AU. {The coordinates of $x$ and left $y$-axis are in units of $R_{\star} $, on the right $y$-axis in AU.}}  
\label{plane_varB}
\end{figure}

Particle scattering off the turbulent magnetic field is greater for stronger turbulence {($\sigma^2 =1.0$)}.  The effect of this is 
shown in the 2D distribution of hitting-points in Fig.~\ref{plane_varB}.  As a result of the smaller mean free path, i.e. $\lambda_\parallel \simeq 1/\sigma^2$ {(see Eq. \ref{lambda})}, for $R_s = 5 R_{\star}$ no hitting-points are detected beyond $\sim 14 R_{\star}$ for $\sigma^2 = 1.0$ {(red dots)}: {EPs scatter back and forth more frequently along a given field line and are more likely to hit the disk close to the injection sphere; such confinement is also favoured by the magnetic field topology, significantly more envelopped around a T ~Tauri star than around the Sun. In the case of weaker turbulence ($\sigma^2 = 0.1 , 0.01$, green and blue dots respectively)  
the scattering is less frequent and a wider spread is found, out to the boundary of the simulation box.} 
The fractions of injected particles hitting the disk are $\sim 99\%,
22\%, 16\%$  for $\sigma^2 = 1.0, 0.1, 0.01$, respectively; the fractions of the particles collapsing back to the star are $\sim 1\%, 60\%, 68\% $, respectively.  In each case, the residual corresponds to the escaped EPs. 

\begin{figure}
	\includegraphics[width=8.5cm]{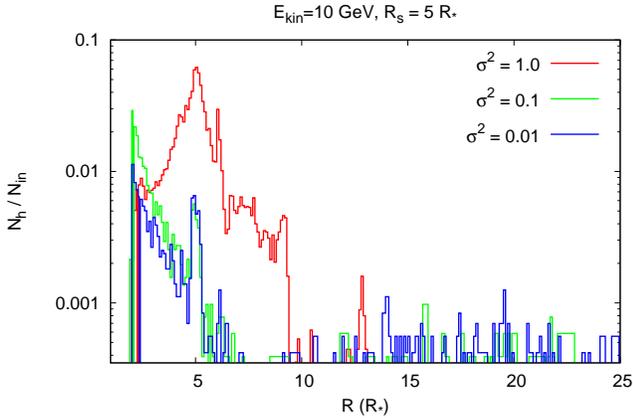}
\caption{Histogram of $N_h / N_{in}$ for $10$ GeV particles as a function of radial distance from the star (in units of $R_{\star}$)  for  $\sigma^2 = 0.01, 0.1, 1.0$ with $L_c = 10^{-5}$ AU. Particles are injected  on a sphere with radius $R_s= 5 R_{\star}$. Note that the y-axis in a log scale. \label{histo_varB}}
\end{figure}

\begin{figure}
	\includegraphics[width=8cm]{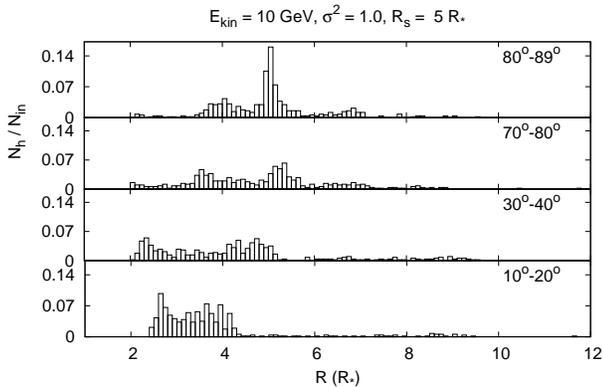}
\caption{Histogram of $N_h / N_{in}$ for $10$ GeV particles as a function of radial distance from the star (in units of $R_{\star}$) for $\sigma^2 = 1.0$. EPs are injected within various rings of co-latitudine $\vartheta$ (from top to bottom $[80^\circ - 89^\circ]$, $[70^\circ - 80^\circ]$, $[30^\circ - 40^\circ]$, $[10^\circ - 20^\circ]$) on a spherical surface with radius $R_s= 5 R_{\star}$. 
}  
\label{histo_comp_varB_lat}
\end{figure}

The more efficient particle confinement ($\lambda_\parallel \ll 10 R_{\star}$) by stronger turbulence is further illustrated by the radial histogram in Fig.~\ref{histo_varB}. For $\sigma ^2 =1.0$ (red), the histogram has a narrow peak at the injection radius $R_s = 5 R_{\star}$ and reaches out to only $\sim 13 R_{\star}$. In contrast, for $\sigma ^2 =0.1, 0.01$ {(green and blue lines respectively)}, hitting points spread out to the box boundary as EPs tend to follow field lines {undisturbed} over greater distances because of the reduced scattering and very small perpendicular diffusion. This also leads to the blue histogram lying somewhat below the green one close to the star.  {Fig.~\ref{histo_comp_varB_lat} shows the radial histogram of equal-number populations of EPs injected within distinct co-latitudinal rings. Particles emitted at low co-latitudes $\vartheta$ ($10^\circ - 20^\circ$, close to the magnetic pole, {lowest panel in Fig.~\ref{histo_comp_varB_lat}}) can hit the disk as a result of enhanced scattering, {unlike the case of weaker turbulence} where EPs injected in regions magnetically connected with the pole {are able to travel farther and} mainly collapse back to the star ({as described earlier}). On the other hand, the two top panels of Fig.~\ref{histo_comp_varB_lat}, corresponding to particles {injected} close to the disk, show that, as a result of shortened mean free path, the hitting points {reach out $10 R_{\star}$}, peaking approximately at the injection radius $R_s = 5 R_{\star}$.   This contrasts with the histograms in Fig.~\ref{histo_comp_se_lat}, where all particle impacts peak at smaller radii and are confined within $R_s = 5 R_{\star}$.  The larger spread for stronger turbulence is a result of increased particle scattering.} 
We emphasize that the histograms presented up to this point can be understood as result of pure scattering of the EPs off the turbulence ($\kappa_\parallel$) with a small contribution of perpendicular diffusion ($\kappa_\perp$) due to the very strong field $B_0$ close to the host star. We will consider in future work larger distances ($>10$ AU), where $\kappa_\perp$ becomes more relevant.
We have also investigated the change of the particle distribution for selected values of $L_c$ (see Appendix \ref{s:Lc}).

\begin{figure}
	\includegraphics[width=9cm]{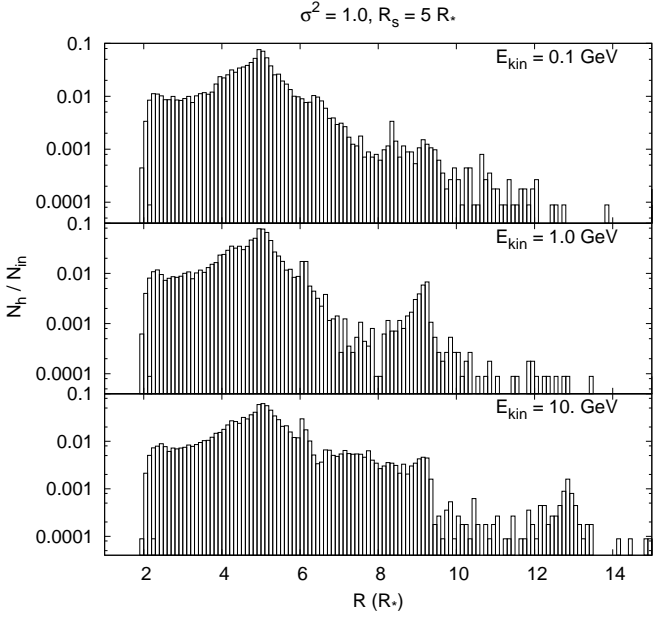}
\caption{Radial histogram of $N_h/N_{in}$ for $E_{kin} = 0.1, 1.0, 10$ GeV protons normalized to the same $N_{in}$ for $\sigma^2 = 1.$, $R_s= 5 R_{\star}$ and $L_c = 10^{-5}$ AU. The coordinates of the $x$-axis are in units of $R_{\star} $. Note that the y-axis is on a log scale.}  
\label{comp_E}
\end{figure}

The stellar EPs flux impinging on the disk is investigated as a function of the energy of the EPs in Fig.~\ref{comp_E} for strong turbulence ($\sigma^2 = 1.0$), where the same number of particles is injected at the same distance ($R_s = 5 R_{\star}$) for three values of energy $E_{kin} = 0.1, 1.0, 10$ GeV. Particles with energy $< 0.1$~GeV do not contribute significantly to the ionization at high mass column \citep{Umebayashi.Nakano:81} and are consequently of less interest here. Higher energy particles (lowest panel) exhibit an appreciably larger radial spread, as expected. \citet{Caramazza.etal:07} found that Orion Nebula Cluster (ONC) T~Tauri X-ray light curves can be entirely explained as a superposition of flares, {with a power-law differential energy distribution $dN/dE \propto E^{-\alpha}$ with a typical index $\alpha \sim 2.2$, where $N$ is the number of flares with energy between $E$ and $E+dE$. We note that measurements of solar energetic particles in the solar wind (cycle 23) have shown an energy dependent correlation between the peak of the proton flux in various energy channels (from $5$ to $200$ MeV) and the intensity of the flares producing the particles \citep{Dierckxsens.etal:15}: such a correlation increases with energy. So it seems reasonable to assume that the power-law index $\alpha$ describes with increasing accuracy the differential spectrum of the EPs produced by the flares as the energy grows; we postpone such an analysis to a separate work.} 
Thus, the panels in Fig.~\ref{comp_E} will undergo an additional normalization  
that reduces the contribution of the $10$ GeV particles to the disk-ionization by a factor $\sim 10^6$ as compared to the $0.1$ GeV particles.
We emphasize that such a spectrum introduces a severe cut-off at high-energy, in contrast with the assumptions of \citet{Feigelson.etal:02} and TD09 who assumed that the EP flux is constant in energy beyond $0.01$ GeV. 

\begin{figure}
	\includegraphics[width=4.5cm]{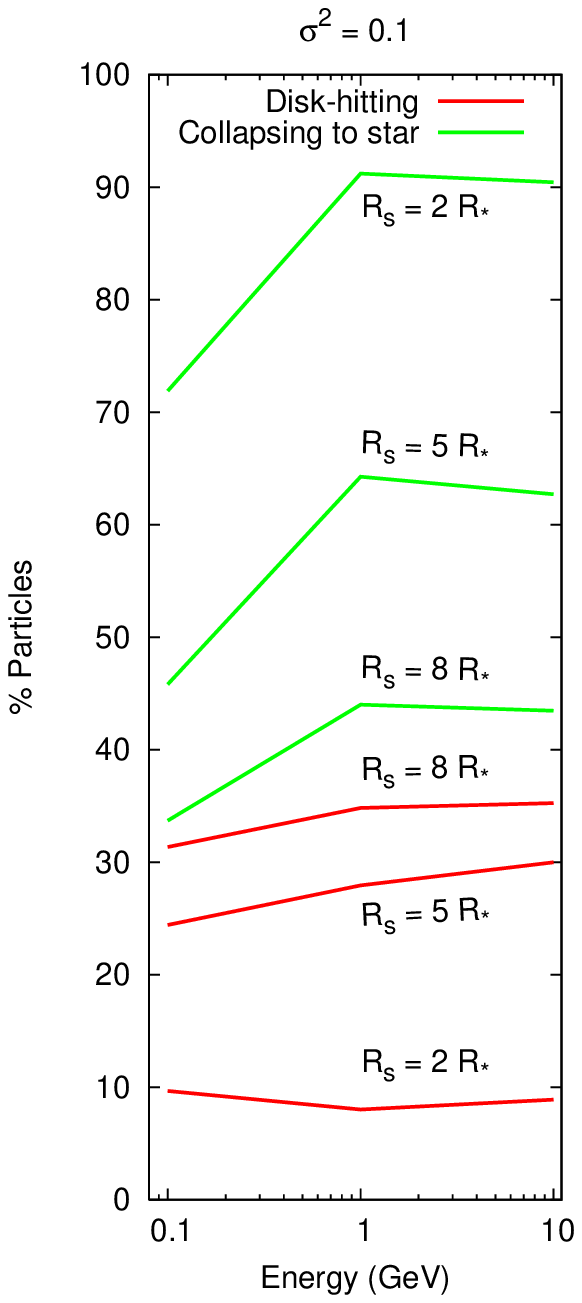} \hspace{-0.7cm}
	\includegraphics[width=4.5cm]{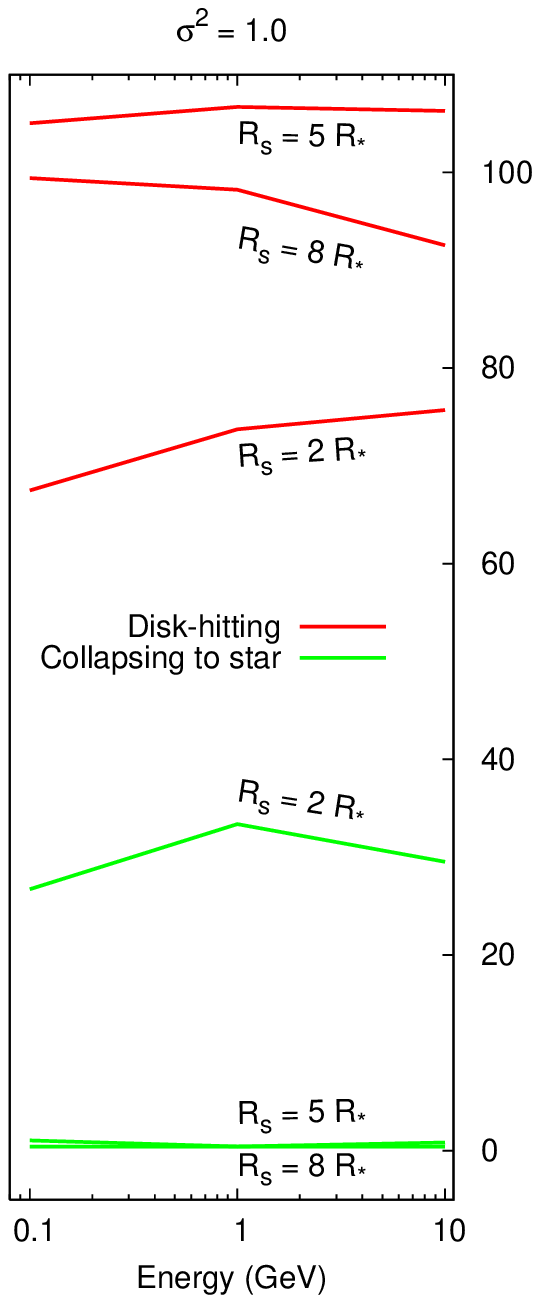}
	\caption{{\bf Left} The fraction of particles hitting the disk (red) and collapsing back to the star (green) with respect to the injected particles for $R_s = 2,5,8 R_{\star}$ as a function of particle kinetic energy (between $0.1$ and $10$ GeV) for $\sigma^2 = 0.1$ and $L_c = 10^{-5}$ AU. {\bf Right} The same as the Left panel except for $\sigma^2 = 1.0$.  \label{Fraction}}
\end{figure}

Figure~\ref{Fraction} shows the fraction of the injected particles that hit the disk and collapse back onto the star for various $R_s$ as a function of particle kinetic energy (between $0.1$ and $10$ GeV) for $\sigma^2 = 0.1$ and $1.0$. As expected, the EPs hitting the disk are relatively unimportant at low $\sigma^2$ (red curves in the left panel), whereas, in the case of high $\sigma^2$, most of the particles are eventually conveyed to the disk. In all cases, the fraction of disk-hitting particles is roughly independent of particle energy

\section{Predicted Ionization Rate}\label{ionrate}

The strength of the ionization of the disk gas arising from the interaction of EPs and the disk surface layers is depicted in Fig.~\ref{ioniz_comp_se} via the ionization rate at columns of 8~g~cm$^{-2}$.  This is directly compared with the analysis of TD09. The two curves from TD09 represent the stellar energetic protons rectilinearly propagating (blue) and the stellar X-rays (purple); both curves scale as $R^{-2}$ and are based on the assumption that the relation between EP flux and ionization rate is independent of the details of the EP energy spectrum. Since the stellar EPs in that study do not follow strongly curved field lines nor are scattered by turbulence, a large fraction of particles emitted by the flaring star will eventually hit the disk at a certain distance, none will collapse back to the star and some will escape along the open field lines, regardless of the value of $E_{kin} > 0.1$~GeV.

The two histograms in Fig.~\ref{ioniz_comp_se} correspond to the proton ionization rate as a function of radial distance from our test particle calculations and represent the disk hitting points of two proton populations of equal number injected at $R_s = 5 R_{\star}$ (red) and $10 R_{\star}$ (green) for $\sigma^2 =1.0$ and $L_c = 10^{-5}$~AU. 
The histograms have been normalized to the TD09 results as follows.  Firstly, we note that the distributions of hitting points are very similar for the different proton energy cases considered (see Fig.~\ref{comp_E}).  To first order, then, our study indicates that the trajectories of {\em ionizing} particles (those with $E\geq 0.1$~GeV) are essentially the same around the peak and independent of energy. We can therefore scale our monoenergetic propagation results in a general way as being representative of all of the ionizing stellar EPs population. A dependence on the EPs spectrum will only reduce the contribution of the higher energy particles with no qualitative change to this diagram; we will include such an effect in a forthcoming work. TD09 assumed a near surface disk 
ionization rate due to energetic stellar particles of $\zeta_{SP}=10^4 \zeta_{CR}\, r_{AU}^{-2}$, where $\zeta_{SP}$ and $\zeta_{CR}$ are the stellar particle and cosmic ray ionization rates, respectively, and $r_{AU}$ is the radial distance from the star in units of AU.  The ionization rates are therefore equal at a radial distance of 100~AU. The cosmic ray ionization rate is $\zeta_{CR}=3.8\times 10^{-18}$~s$^{-1}$ based on Eqn.~2 of TD09 for a mass column depth of 8~g~cm$^{-2}$ and a characteristic cosmic ray absorption depth column of 96~g~cm$^{-2}$.  The ionization rate from our test stellar particle simulations relative to the TD09 expression is then 
\begin{equation}
\zeta_{TSP}(r) 
= 3.8\times 10^{-18}\, \left(\frac{100}{r_{AU}}\right)^2\, \frac{n_h(r)}{n_l(r)} \; {\rm s}^{-1},
\end{equation}
where $n_h(r)$ is the actual number of disk-hitting protons per cm$^{-2}$ at radial distance $r$ from our simulations, and $n_l(r)$ is the expected flux were all the injected particles to have traveled outwards isotropically and linearly from a central point. Applying the TD09 projection factor accounting for the shallow angle of impact of stellar protons on the disk due to its vertical flare as a function of radial distance by an angle of approximately 0.1~radian, the latter quantity is simply 
\begin{equation}
n_l(r)=\frac{0.1\,N_{in}}{4\pi r^2}.
\end{equation}
The final ionization rate is then 
\begin{equation}
\zeta_{TSP}=3.8\times 10^{-18} \left(\frac{100}{r_{AU}}\right)^2\, \frac{4\pi r^2}{0.1\,N_{in}}\, n_h(r) 
\end{equation}
which is illustrated in Figure~\ref{ioniz_comp_se}.

\section{Discussion}\label{discuss}

\begin{figure}
	\includegraphics[width=8.5cm]{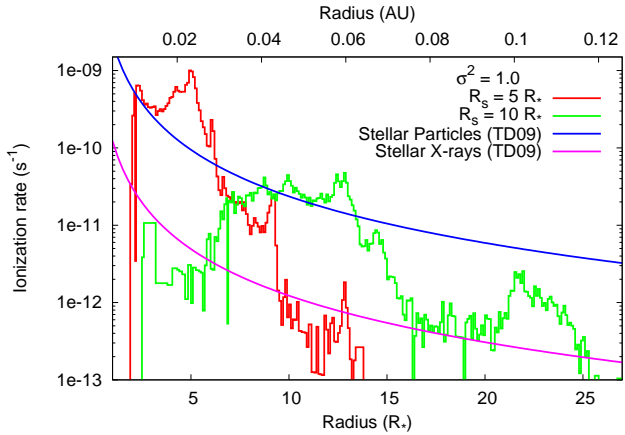}
\caption{Ionization rate at columns of 8~g cm$^{-2}$. Radial histograms represent the contribution of EPs injected at $R_s = 5R_{\star}$ (red) and $10 R_{\star}$ (green) for $\sigma^2 = 1.0$ and $L_c = 10^{-5}$ AU. The two solid lines decreasing as $r^{-2}$ are reproduced from TD09 and show the contribution to the disk-ionization from radially propagating stellar EPs (blue) and from stellar X-rays (purple).  \label{ioniz_comp_se}}
\end{figure}

Figure~\ref{ioniz_comp_se} shows that the fate of EPs at a certain distance from the star in a realistic $B$-field differs markedly from the assumption of rectilinear propagation.
Not only does the large-scale magnetic field and the winding of the field lines around the star hamper linear, radial propagation, but also the strong magnetic turbulence reduces $\lambda_\parallel$, confining the motion to a small radial interval around $R_s$, in contrast with the large $r^{-2}$ spread assumed in TD09. As a result of such a two-fold effect, the ionization of the disk is dominated 
by stellar EPs in a mottled fashion---only in localized regions of the disk, and in our simulations within a few $R_{\star} $ from the region of EP injection. The large-scale ionization of most of the disk will likely be dominated by radially propagating $X$-rays, whose ionization rate decays as $r^{-2}$.
On the other hand, it is also likely that EPs can locally dominate over stellar $X$-rays in regions well beyond our present simulation box, such that the true ionization structure at any given time could be quite spatially inhomogenous.

We have shown in Fig.~\ref{histo_varB} that EPs spread to larger distances if the turbulence is weaker, and that this leads to a narrower but closer-in region of EP-dominated ionization of the disk. 

The resonance condition of the particle gyroradius with a given inertial scale is satisfied to good approximation here. The EPs that leave the simulation box along open field lines experience a decrease of the large-scale magnetic field from the injection point to the box boundary ($\simeq 25  R_{\star}$) of a few orders of magnitude; the corresponding increase of $r_g$ during outward motion might invalidate the resonance condition. However, our simulations have shown no statistical variation in the 2D distribution and in the radial histogram of the disk-hitting points by extending the inertial range of the turbulence power spectrum by one decade to both smaller and greater wavenumbers (see Sect.~\ref{results}). 

Although we have used a steady magnetic field configuration as a background for EP propagation, such a field structure can be considerably different from the one left behind a propagating CME preceded by a shock: many field lines are compressed, and reconnection can occur at various places. The fractions of particles collapsing back to the star, and also leaving the simulation box, might change during transient events, e.g., CMEs, that disrupt the steady structure of the magnetic field behind them, and advect EPs away from the star. In this work, we neglect such a time-dependence of the magnetic field configuration; it is plausibly of second order importance in considering the EP ionization rate, although a time-dependent study would be well-motivated.

Evidence of high fluxes of stellar EPs on protoplanetary disks that induce spallation reactions and produce short-lived nuclei has been reported by \citet{Ceccarelli.etal:14} based on {\it Herschel} observations of the young star OMC-2 FIR~4. The estimated proton flux for particles with energy $E \geq 10$~MeV is of the order of $10^{11}$--$10^{12}$~protons~cm$^{-2}$~s$^{-1}$ at 1~AU.  This is four to five orders of magnitude larger than the estimate of \citet{Feigelson.etal:02} for typical T~Tauri star and a challengingly high flux: inverse square geometric dilution with radial distance for an average proton energy of a few tens of MeV implies a proton ``luminosity" similar to or greater than the total stellar luminosity.
If the estimate proves to be close to correct, it would provide direct proof that proton trajectories in the circumstellar environment of a T~Tauri star are highly non-rectilinear and channeled toward the disk.

Radio imaging supports high fluxes of energetic electrons from T Tauri stars. The radio-polarization in the bi-polar outflow of the binary T Tauri stars has been explained \citep[ADDED REF]{Ray.etal:97} with a power law distribution of mildly relativistic electrons (Lorentz $\gamma \sim 2$--3) possibly accelerated by a shock within the outflow, rather than in a coronal flare. \citet[ADDED REF]{AMI.Ainsworth.etal:12} reported variability in the radio luminosity of two sources (L1551 IRS 5 and Serpens SMM 1) superposed to thermal bremsstrahlung continuum emission; however, no firm constrains was found on the electron energy  distribution of the transient.

In this context, the solution of the transport equation by \citet{Rodgers-Lee.etal:17} inferred an $r^{-1}$ dependence of energetic proton intensity with radius, or steeper depending on the assumed diffusion coefficient, encoding all the turbulence details in a diffusion coefficient uniform throughout. The ionization rate they deduced follows the expression in \citet[Eqn.~23]{Umebayashi.Nakano:81} for a selected initial EP energy ($3$ GeV) that depends on the energy losses as particles traverse the disk and drops steeper than $r^{-3.5}$.  While computation limitations restrict our present study to the very inner disk region, one common finding between our study and \citet{Rodgers-Lee.etal:17} is that the effect of stellar EP ionization is largely confined by magnetic turbulence to the region where EPs are injected. However, one unique prediction of our study is that ionization will likely be spatially non-uniform and ``mottled'' to some extent owing to stellar magnetic field and wind inhomogeneities and their effects on particle transport.  

Transient accretion onto the star through equatorial tongues and high-latitude funnel flows generated by the MHD instabilities in 3D simulations \citep[ADDED REF]{Kulkarni.Romanova:08} at the disk--magnetosphere boundary have been connected to non-periodic bursts in light curves of young stars \citep[see for ONC][ADDED REF]{Herbst.etal:02}. The high density of the tongues, comparable to the density of the disk, can prevent EPs originating from coronal flares or from shocks close to the star from reaching further out or escaping. On the other hand, funnels break the magnetic topology wrapped around the star, opening channels for EPs to escape non-adiabatically. A proper assessment of the disk ionization requires a separate set of 3D-MHD simulations. We note that, since the pressure of EPs is negligible compared to the thermal gas pressure, EPs are not expected to affect the development of Rayleigh-Taylor instabilities, unlike numerically found, for instance, in the shocked layer of supernova remnants \citep[ADDED REF]{Fraschetti.etal:10}. 

In this work, we adopted a representative rotation period $4$ days. The primary influence of the rotation period is through the azimuthal wrapping of the magnetic field, with faster rotation leading to a stronger circumstellar field.  A different choice within a few days of that adopted, e.g.\ the $6.53$ day period of V2129~Oph, would not have led to a significant change in the distribution of EP disk-hitting points. In general, for increasing rotation rate we expect that EPs would spend more time in the transversal direction and less time in the radial direction, and with stronger confinement in the ionizing streams and spots.

There are four time scales that could affect the ionization pattern caused by EPs. (1) The magnetic field on the stellar surface will reverse polarity within a stellar cycle ($\sim 11$ years for the Sun and presumably comparable for a T~Tauri star). (2) The surface magnetic field pattern is liable to change on timescales of days to weeks, analogous to that on the Sun, as new magnetic field emerges and older field is dissipated or subducted. Random plasma motion on the stellar surface could also generate fluctuations that would be carried outward by the wind and are therefore quasi-static on a time scale much greater than a few hours. The latter provides a simple model of the highly anisotropic interplanetary magnetic turbulence in the solar wind \citep[ADDED REF]{Giacalone.etal:06}.
(3) In the stellar wind and very close to the star, within a few $R_\star$ or the Alfv\'en radius, the EP-generated ionization patterns will be dragged by stellar rotation relative to the disk. (4) Further out, the unperturbed field advected by the stellar wind will dominate by stretching the patterns radially on the disk surface. The EP propagation time, within a few hours in most cases, is much shorter than all these time scales.

The persistence of the EP ionization pattern, and hence its {\em observational} relevance, can be estimated by comparing the EP propagation time scale with the time scale of recombination of free electrons within the disk  \citep[ADDED REF]{Ilgner.Nelson:06c}. At distances $< 1$~AU from the star (in this work we consider $R < 0.2$ AU), the recombination time scale is of the order of $\sim 1$ day or less and is shorter than the typical time between large stellar $X$-ray flaring outbursts; the latter is of the order of $\sim 1$ week with a typical flare duration of a few hours, as observed during the {\it Chandra} Orion Ultradeep Project (COUP) and reported by \citet[ADDED REF]{Wolk.etal:05}. Further out (distances $> 2$ AU), the recombination time scale is comparable to $\sim 1$ week. The reason for the difference is that close to the star the recombination is dominated by molecular ions, whereas at larger distances by heavy metals and hence is much slower\footnote{The recombination rate can be faster by orders of magnitude in the presence of dust grains whose growth and settling was examined by \citet[ADDED REF]{Ilgner.Nelson:06a}. In addition, the concentration of metals in the outer disk will also be affected as they settle onto the grains and cannot contribute to recombination. Here we do not examine the effect of grains.}. Therefore, within $0.2$ AU from the star, the inhomogeneous ionization pattern might disappear via recombination faster than the local disk orbital period, making a direct observation of ionization spots challenging; in addition, high spatial resolution would be needed (the linear size of the spots is $< 0.1$ AU). However, this conclusion does not change the main result that the ionization is only locally, and not globally, supported by EPs as compared with $X$-rays from coronal flares. Further out, the recombination time is much longer and the ionized regions are broader, potentially allowing persistent spots to be observed. The disk evolutionary time ($\sim$ Myr) is much longer than all the aforementioned time scales, and so the fact that the ionization pattern is inhomogeneous is unlikely to have specific direct consequences as compared with a uniform ionization. 
\citet[ADDED REF]{Winters.etal:03} report a chaotic development of turbulence as a result of MRI that, arguably, should lead to a very short convection time, perhaps shorter than the recombination. In this case the mottled ionization would be a relevant contribution to the disk viscosity.

We note that the innermost region of the disk ($R < 0.1 $ AU) is sufficiently hot \citep{Gammie:96} to ionize heavy atoms (such as sodium or potassium): such thermal ionization is sufficient to couple the flow to the magnetic field, thereby affecting MRI turbulence, even without EP ionization. The magnetic field produced in radiative 3D-MHD simulations in \citep[ADDED REF]{Hirose:15} might contribute the total field and influence the EPs propagation.

Current observations are not sufficient to assess whether EPs from stellar flares or travelling shock waves dilute continuously outward or rather are localized in spots in such a way by the topology of the turbulent magnetic field. However, 
observations of molecular species sensitive to the local ionization rate with the high spatial resolution of the Atacama Large Millimeter Array telescope should provide some constraints and help disentangle the contributions of stellar X-rays and EPs to disk ionization rates, as well as the influence of any variability in photon or particle fluxes \citep[e.g.][]{Cleeves.etal:15,Cleeves.etal:17}.

Finally, we note two salient aspects of our study that should be improved upon.  Firstly, due to both limitations in computing capacity and the size of the simulation box that could be treated in our MHD wind model, the present study is limited to the inner disk not far from the parent star.  In order to present predictions for T~Tauri star EP ionization for regions of a disk that are more interesting for planet formation and testable by spatially-resolved observations, the test particle calculations need to extend to much larger radial distances.  While the stellar wind dominates the magnetic and outflow structure within the bounds of our model, further out a disk wind structure is expected to provide the dominant ambient magnetic field and plasma flow \citep[e.g.][]{Bai:17}.  There will also be a complex and potentially very turbulent interface region in which both the stellar wind and disk wind present similar ambient pressures. While including both stellar and disk winds in particle transport models such as that presented here will be challenging, this is probably the only route leading to a realistic and rigorous investigation of disk ionization by stellar EPs and cosmic rays. 

Secondly, the true EP energy spectrum for a T~Tauri star environment remains unknown. 
Lacking direct information, the EP spectrum has generally been simplified by existing studies to a uniform spectrum in energy or a monochromatic EP flux, or some scaling of solar EP distributions has been adopted. T~Tauri flares are orders of magnitude more powerful than the most energetic solar flares observed to date and their EP production can presently only be speculatively extrapolated from observed solar events. The true EP flux and energy distribution will remain a large uncertainty in protoplanetary disk ionization studies until either models of their acceleration can be sufficiently improved as to be reasonably applied to T~Tauri stars, or radio bursts produced by coronal mass ejections will enable constraints to be placed on the acceleration process.

\section{Conclusions}\label{conclusions}

We have performed test-particle simulations to propagate stellar energetic particles through a realistic and turbulent magnetic field of a young solar mass T Tauri star to investigate the effect on the ionization of the inner protoplanetary disk. We have compared the ionization rate in the disk due to stellar EPs and to the steady flux of stellar X-rays.  Since the tangled and turbulent magnetic field lines hamper any steady outflow of ionizing energetic particles, we find that the large-scale ionization of much of the disk will likely be dominated by radially propagating $X$-rays, whose ionization rate decays as $r^{-2}$. However, by mimicking the steady injection of particles from flares or shock waves into the circumstellar medium by spherically injecting particles at various radii, we find that regions of the disk emerge that are predominantly ionized by stellar energetic particles.

The channelling of particles toward the disk is more efficient for stronger turbulence, and a result of the parallel diffusion only since the perpendicular diffusion is small in the strong unperturbed magnetic field close to the star. However, we expect the role of perpendicular transport to grow at distances of $\sim 1$ AU or greater from the star. We speculate that EP-induced ionization spots could extend out to large distances from the star, as particles are efficiently carried by travelling shocks as observed in the solar wind. 

A full understanding of protoplanetary disk ionization will require more complete knowledge of the EP flux and energy spectrum produced by T~Tauri star magnetic activity, and treatment of the combined stellar and disk winds and magnetic fields.

\acknowledgments

We gratefully acknowledge the careful reading and the helpful comments of the referee. We extend warm thanks to Xuening Bai for insightful discussion and for comments that enabled us to improve and clarify the manuscript. FF also acknowledges continuous discussions with J. Kota, J. Giacalone and J. R. Jokipii. Support for this work was provided by the National Aeronautics and Space Administration through Chandra Award Number $TM6-17001B$ issued by the {\it Chandra X-ray Center} (CXC), which is operated by the Smithsonian Astrophysical Observatory for and on behalf of the National Aeronautics Space Administration under contract NAS8-03060. The work of FF was supported, in part, also by NASA under Grant  NNX15AJ71G. OC is supported by NASA LWS grant NNX16AC11G. JJD was funded by NASA contract NAS8-03060 to the CXC and thanks the Director, Belinda Wilkes, for continuing advice and support. Resources supporting this work were partially provided by the NASA High-End Computing (HEC) Program through the NASA Advanced Supercomputing (NAS) Division at Ames Research Center. This work benefited from technical support by the computer cluster team at the Department of Planetary Sciences at University of Arizona.

\appendix

\section{Distribution of the hitting-points for different values of $L_c$}\label{s:Lc}

\begin{figure}
	\includegraphics[width=8.5cm]{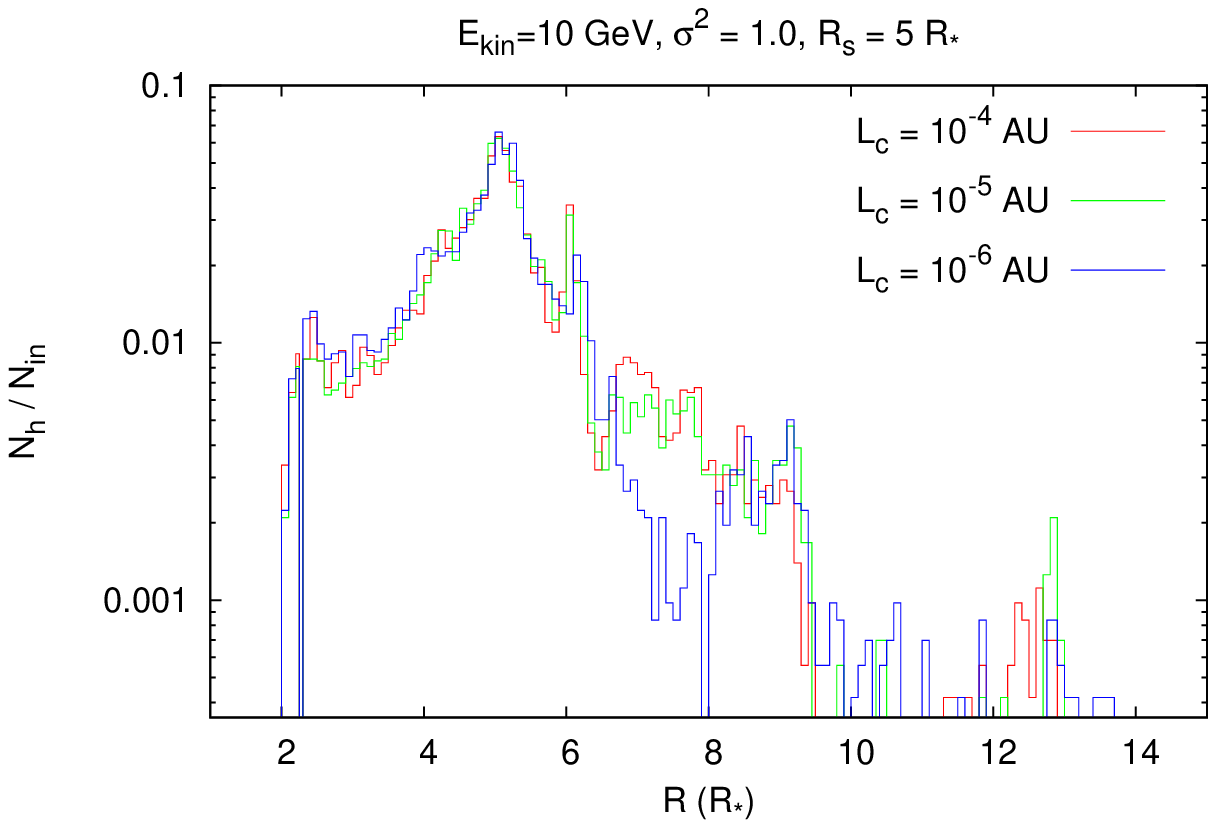}
	\includegraphics[width=8.5cm]{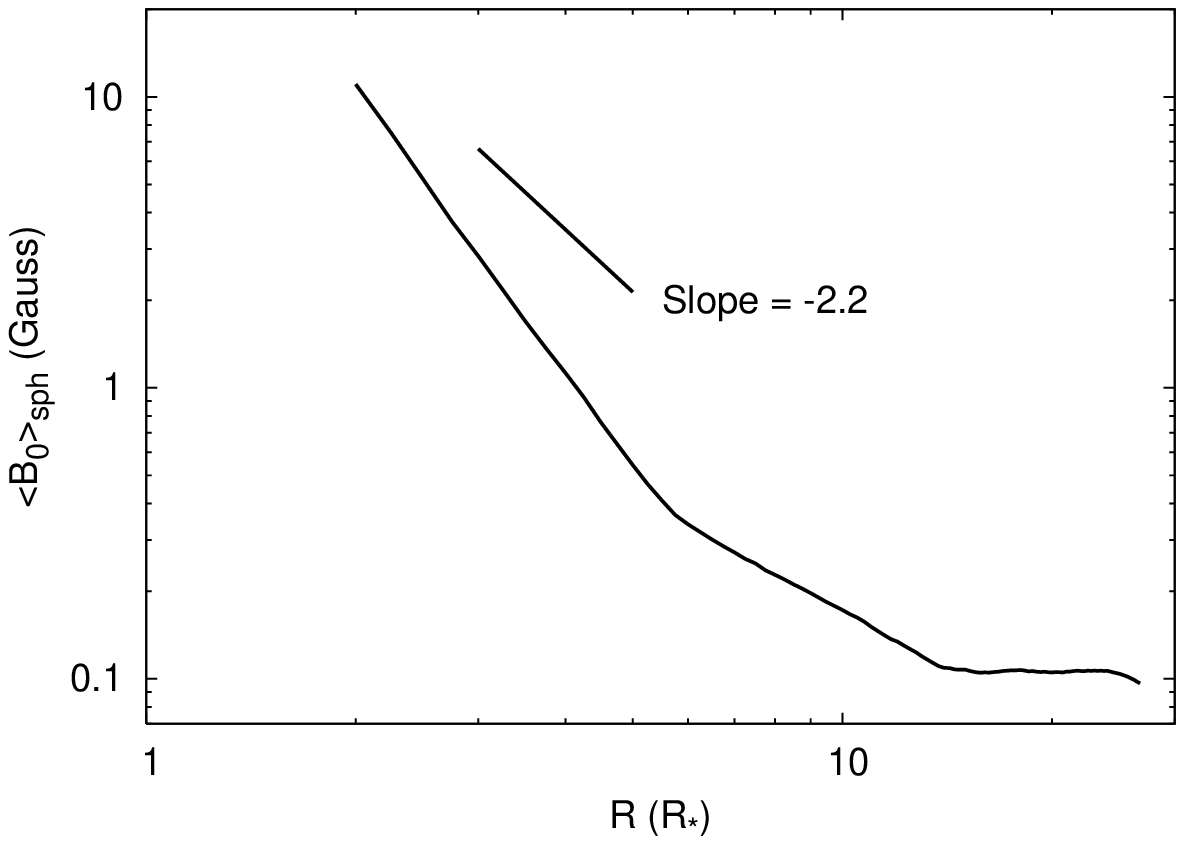}
	\caption{{\bf Left}: Radial histogram of hitting points for $E_{kin} = 10$ GeV kinetic energy protons on the disk plane normalized to the number of injected particles for three values of $L_c = 10^{-4}, 10^{-5}, 10^{-6}$ AU and $\sigma^2 = 1.0$. The coordinates of the $x$-axis are in units of $R_{\star}$. Protons are injected on a sphere with radius $R_s = 5 R_{\star} $. {\bf Right}: Radial dependence of the spherical average of the magnitude of the unperturbed magnetic field $B_0 (\mathbf x)$ for $R > 2R_{\star}$ compared with a power law with index $-2.2$ (see text). \label{histo_comp_Lc}}
\end{figure}

Figure \ref{histo_comp_Lc}, Left panel, shows the radial histogram of the hitting points on the plane of the disk (the XY plane) for $E_{kin} = 10$ GeV, $R = 5 R_{\star}$, $\sigma^2 = 1.0$ and selected values of $L_c$. The chosen range of $L_c$ ($10^{-6} - 10^{-4}$ AU) is such that the resonance condition for the protons in the kinetic energy range $0.1$ to $10$ GeV applies to propagation within most of the simulation box. The three curves show a significant overlap except in the range $6.5 - 8 R_{\star}$.  At these distances, the number of EPs drops by a factor $\sim 5$ for lower $L_c$ ($\lambda_\parallel \simeq L_c^{2/3}$ from Eq.~\ref{lambda}). We argue that this results from choosing a single realization of the turbulence $\delta {\bf B}$ and not by averaging over an ensemble of turbulence realizations; such an average is expected to smear out the discrepancy between the three curves in that narrow region.  We conclude that, with the resolution of the spatial grid used to calculate the total magnetic field, different values of $L_c$ do not affect significantly the distribution of disk-hitting particles in the circumstellar medium of a T~ Tauri star.

\section{Turbulence power}\label{s:sigma}

The assumption of a uniform $\sigma^2$ throughout the simulation box entails a spatial variation of the amplitude of the fluctuation $|\delta {\bf B}|$ (calculated as sum of plane waves with random orientation, polarization and phase) beside the variation of the phase. Figure~\ref{histo_comp_Lc}, Right panel, shows the radial dependence of the spherical average of the unperturbed field ${\bf B}_0 (\mathbf x)$ for $R > 2R_{\star}$ produced by our MHD simulations. Measurements of the amplitude squared of the magnetic fluctuations of the solar wind at low-latitudes by {\it Helios} between $0.3$ and 1 AU and in the polar regions by {\it Ulysses} out to $4$ AU \citep[see Fig. 4.8 therein]{Horbury.Tsurutani:01} yield a power law  dependence on heliocentric distance with index $-2.2$, as illustrated in Fig.~\ref{histo_comp_Lc}, Right panel. Such observations support our assumption that $|\delta B|$ falls off with the distance from the star, thus it seems reasonable to simplistically assume a uniform $\sigma^2$.

The prescription used to calculate the magnetic fluctuation as a sum of plane waves with a space-dependent amplitude requires enforcing separately that $\nabla \cdot \delta {\bf B} ({\mathbf x}) = 0$, beside the condition $\nabla \cdot {\bf B}_0 ({\mathbf x}) = 0$ already satisfied by the MHD grid with an assigned numerical precision. A method to impose the solenoidal constraint on a spatially varying turbulence with uniform $\sigma^2$ is not known at present, even for the case of the Parker Archimedean spiral field in the solar wind. For the simulations presented here, the correlation length of the turbulence, $L_c$, is smaller than the cell size ($0.36 R_\star$) where $B_0$ is uniform; thus, in our simulations $\delta B$ has cell-by-cell uniform amplitude and the approach in \cite{Fraschetti.Giacalone:12}  applies with no change. In addition, we show with the following qualitative argument that even for $L_c > 0.36 R_\star$ the solenoidal condition on $\delta B$ is satisfied within the numerical accuracy of the MHD grid. Noting that the divergence $\nabla \cdot \delta {\bf B} $ can be written as $\nabla \cdot (|\delta {\bf B}| \hat{\delta {\bf B}}) $, where $\hat{\delta {\bf B}}$ indicates the unit vector of $\delta {\bf B}$, we have 
\begin{equation}
\nabla \cdot (|\delta {\bf B}| \hat{\delta {\bf B}}) = |\delta {\bf B}| \nabla \cdot \hat{\delta {\bf B}} +  (\nabla |\delta {\bf B}|) \cdot \hat{\delta {\bf B}}. 
\end{equation}
For turbulence with space-independent amplitude, $\nabla \cdot \hat{\delta {\bf B}} = 0$ follows from the definition of $\delta {\bf B}$ \citep{Giacalone.Jokipii:99}. The remaining condition, 
\begin{equation}
\sigma (\nabla {\bf B}_0) \cdot \hat{\delta {\bf B}} \leq \nabla \cdot {\bf B}_0, 
\end{equation}
is qualitatively satisfied by noting that 
\begin{equation}
\nabla \cdot {\bf B}_0 \sim B_0 /L, 
\end{equation}
where
\begin{equation}
L = B_0/|\nabla B_0|, 
\end{equation}
and 
\begin{equation}
\sigma (\nabla {\bf B}_0) \cdot \hat{\delta {\bf B}} \sim \sigma B_0 /L \leq B_0 /L. 
\end{equation}

\section{Distribution of the hitting-points for different disk structure}\label{s:disk}

\begin{figure}
	\includegraphics[width=10.cm]{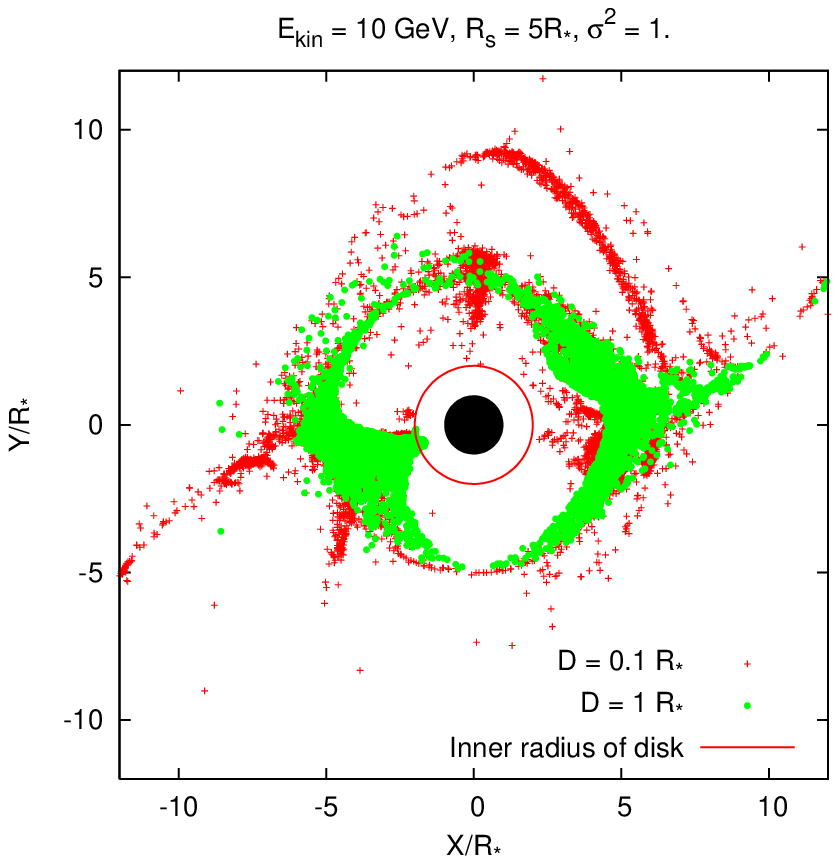} \hspace{-3cm}
	\includegraphics[width=10.cm]{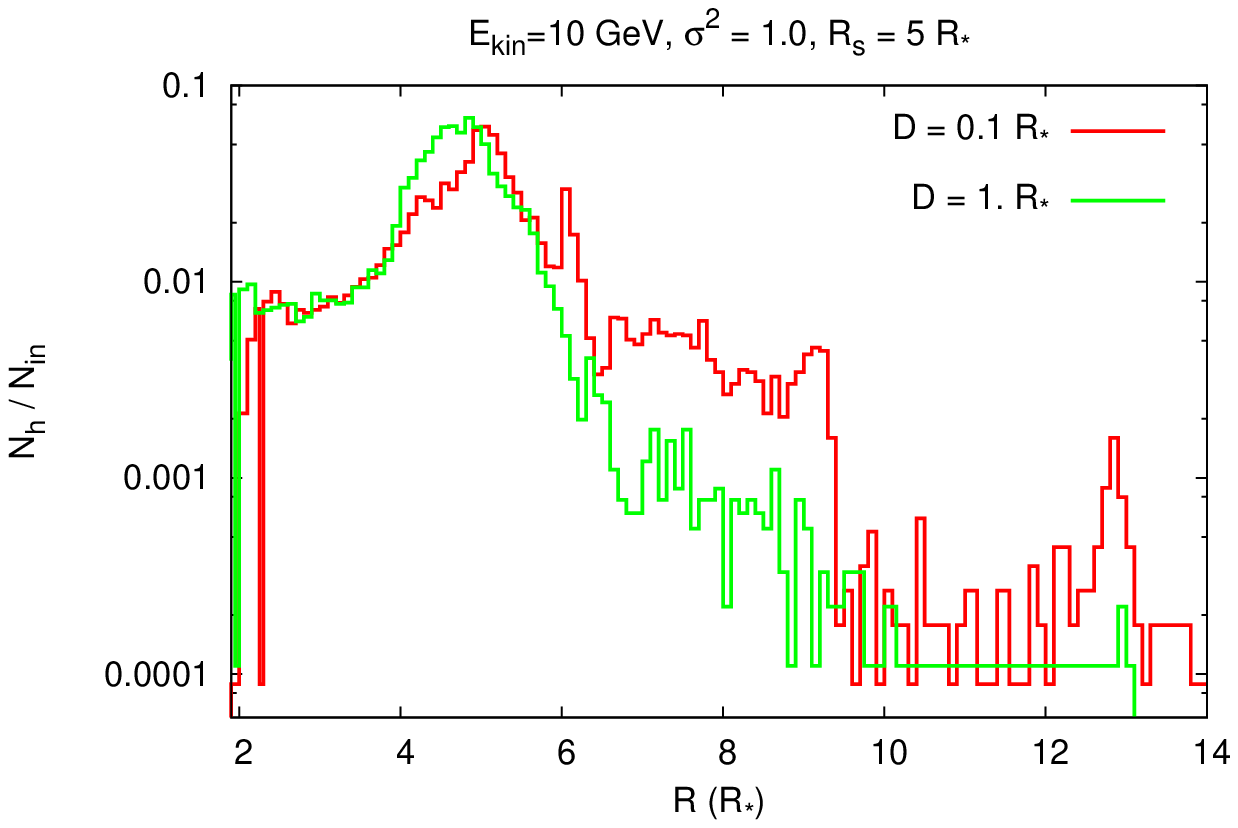}
	\caption{{\bf Left}: Coordinates of the hitting points for 10~GeV kinetic energy protons on the disk plane for two selected values of disk semi-thickness $D = 0.1 R_{\star}$ (in red) and $D = R_{\star}$ (in green); here, $R_s = 5 R_{\star}$, $L_c = 10^{-5}$~AU and $\sigma^2 = 1.0$. The red circle with radius $2 R_{\star}$ marks the innermost radius of the disk. The $x$ and $y$ axes are in units of $R_{\star}$. {\bf Right}: Radial histogram of hitting points on the disk plane normalized to the number of injected particles for the two cases in the left panel. \label{histo_comp_D}}
\end{figure}

\begin{figure}
	\includegraphics[width=10.cm]{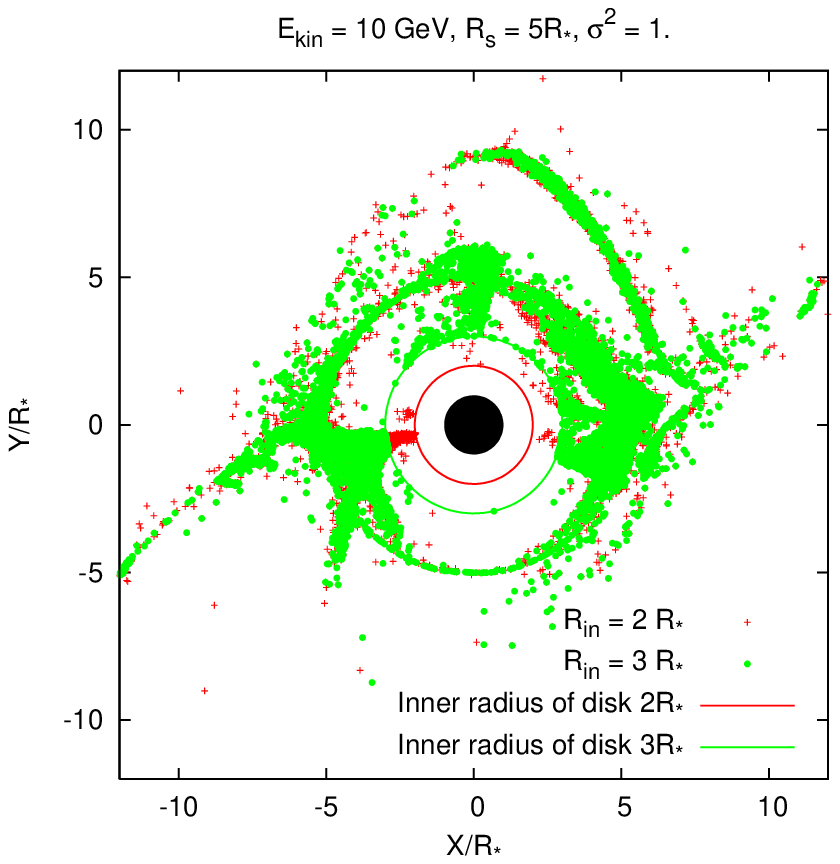} \hspace{-3cm}
	\includegraphics[width=10.cm]{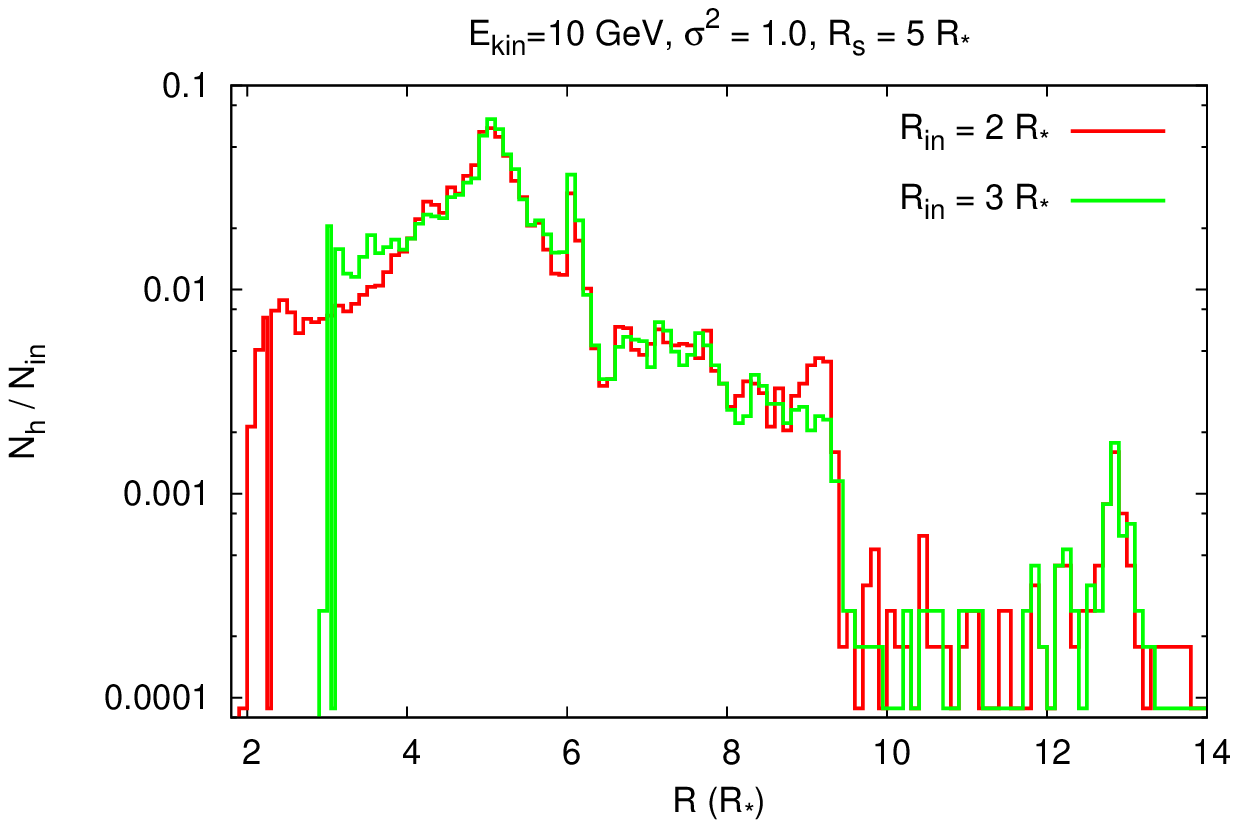}
	\caption{{\bf Left}: Coordinates of the hitting points for 10~GeV kinetic energy protons on the disk plane for two selected values of innermost disk radius $R_{in} = 2 R_{\star}$ (in red) and $R_{in} = 3 R_{\star}$ (in green); here, $R_s = 5 R_{\star}$, $L_c = 10^{-5}$~AU and $\sigma^2 = 1.0$. The $x$ and $y$ axes are in units of $R_{\star}$. {\bf Right}: Radial histogram of hitting points on the disk plane normalized to the number of injected particles for the two cases in the left panel. \label{histo_comp_Rin}}
\end{figure}

Figure \ref{histo_comp_D} compares the distribution of the hitting points on the disk for two distinct values of disk semi-thickness, $D$, uniform throughout the disk (and $\sigma^2 =1$): $D = 0.1 R_{\star}$ and $ R_{\star}$. As expected, the confinement of the hitting points on the disk is even more enhanced because particles travel shorter distances in the latitudinal direction before hitting the much thicker disk. Figure \ref{histo_comp_Rin} compares the cases of different sizes of the magnetospheric cavity, by adopting two values of the disk innermost radius ($2R_\star$ and $3R_\star$). 

\def \apss{{\it Astrophys.\ Sp.\ Sci.}}
\def \aj{{\it AJ}}
\def \apj{{\it ApJ}}
\def \apjl{{\it ApJL}}
\def \apjs{{\it ApJS}}
\def \araa{{\it Ann. Rev. A \& A}}
\def \prc{{\it Phys.\ Rev.\ C}}
\def \aap{{\it A\&A}}
\def \aaps{{\it A\&ASS}}
\def \mnras{{\it MNRAS}}
\def \physscr{{\it Phys.\ Scripta}}
\def \pasp{{\it Publ.\ Astron.\ Soc.\ Pac.}}
\def \gca{{\it Geochim. Cosmochim.\ Act.}}
\def \nat{{\it Nature}}
\def \solphys{{\it Sol.\ Phys.}}

\bibliographystyle{aasjournal}
\bibliography{ffraschetti}



\end{document}